\documentclass{article}
\usepackage{graphicx} 
\usepackage{geometry}
\usepackage{hyperref}
\usepackage{amssymb}
\usepackage{float}
\usepackage{booktabs}
\usepackage{cite}
\usepackage{amsmath}
\usepackage{subcaption}
\usepackage{algorithm}
\usepackage{cleveref}
\usepackage{algpseudocode}
\usepackage[table]{xcolor} 
\usepackage{wrapfig}
\usepackage{authblk}

\title{Product Manifold Machine Learning for Physics}
\author[1,2*]{Nathaniel S. Woodward}
\author[1,2]{Sang Eon Park}
\author[1,2,3]{Gaia Grosso}
\author[4]{Jeffrey Krupa} 
\author[1,2]{Philip Harris}
\affil[1]{NSF AI Institute for Artificial Intelligence and Fundamental Interactions}
\affil[2]{MIT Laboratory for Nuclear Science, Cambridge, MA}
\affil[3]{School of Engineering and Applied Sciences, Harvard University, Cambridge, MA}
\affil[4]{SLAC National Accelerator Laboratory, Menlo Park, CA}
\affil[*]{Corresponding author: \texttt{nswood@mit.edu}}
\date{December 2024}

\begin{document}
\maketitle
\begin{abstract}
Physical data are representations of the fundamental laws governing the Universe, hiding complex compositional structures often well captured by hierarchical graphs. Hyperbolic spaces are endowed with a non-Euclidean geometry that naturally embeds those structures.
To leverage the benefits of non-Euclidean geometries in representing natural data we develop machine learning on \(\mathcal P \mathcal M\) spaces, Cartesian products of constant curvature Riemannian manifolds. As a use case we consider the classification of ``jets", sprays of hadrons and other subatomic particles produced by the hadronization of quarks and gluons in collider experiments. 
We compare the performance of $\mathcal P \mathcal M$-MLP and $\mathcal P \mathcal M$-Transformer models across several possible representations. 
Our experiments show that \(\mathcal P \mathcal M\) representations generally perform equal or better to fully Euclidean models of similar size, with the most significant gains found for highly hierarchical jets and small models. 
We discover significant correlation between the degree of hierarchical structure at a per-jet level and classification performance with the $\mathcal P \mathcal M$-Transformer in top tagging benchmarks. 
This is a promising result highlighting a potential direction for further improving machine learning model performance through tailoring geometric representation at a per-sample level in hierarchical datasets. 
These results reinforce the view of geometric representation as a key parameter in maximizing both performance and efficiency of machine learning on natural data.
\end{abstract}

\newpage 

\tableofcontents

\newpage
\section{Introduction}
\label{sec:intro}
Natural data, from biological systems to the vast expanse of the cosmos, frequently exhibit intricate hierarchical structures. These hierarchies, reflecting multi-scale and compositional relationships, are fundamental to understanding complex systems. Representing such data poses significant challenges in machine learning, where traditional Euclidean spaces often fall short of capturing the underlying relationships. Recent advances have highlighted the utility of non-Euclidean geometries, particularly hyperbolic spaces, in embedding hierarchical data due to their exponential growth properties, which naturally align with tree-like or graph-based data structures \cite{linial1995geometry}. This paradigm shift is supported by studies demonstrating the benefits of hyperbolic embeddings in diverse applications, including natural language processing, social network analysis, and bioinformatics~\cite{nickel2017poincare,tifrea2018poincar}. These findings suggest that geometric representations are not merely a mathematical convenience but a key determinant of model performance and efficiency.

In this paper, we bridge these developments by employing a machine learning framework that leverages \(\mathcal P \mathcal M\) spaces—Cartesian products of constant curvature Riemannian manifolds\cite{bachmann_2020_constant}. These spaces offer a flexible way to represent complex hierarchical data by combining the strengths of different geometries, including hyperbolic and Euclidean components.

In particle physics, the study of jets offers a compelling case for exploring these geometries. Jets are collimated sprays of hadrons and other particles produced during the hadronization of quarks and gluons, typically observed in high-energy collider experiments. These particle cascades are inherently hierarchical, arising from sequential branchings governed by quantum chromodynamics (QCD). Understanding jets is central to probing the fundamental interactions in the Standard Model and beyond, as they play a critical role in identifying particles such as the Higgs boson and exploring phenomena like dark matter production. Consequently, accurate and efficient jet classification is a longstanding challenge, with implications for both theoretical and experimental physics.

Traditional approaches to jet classification have relied heavily on features derived from physical intuition, such as jet substructure observables~\cite{butter2019gan}. However, the increasing availability of high-fidelity simulated data and advances in machine learning have opened new avenues for directly leveraging the full richness of jet data. In particular, the hierarchical nature of jet formation motivates the use of graph-based and geometric learning frameworks, which can capture the underlying structure more effectively than conventional methods.

\paragraph{Our Approach}Hyperbolic spaces provide a natural representation for the hierarchical structure of jets due to their exponential growth in volume with distance\footnote{This is understood through comparing the distortion of tree graphs embeddings in different geometric spaces. Euclidean spaces are unable to achieve comparably low-distortion embeddings compared to hyperbolic spaces \cite{linial1995geometry,peng2021hyperbolicdeepneuralnetworks}.}, however, it is unclear if this representation is optimal for all features of particle jets and all tasks for their analysis. 
Distinct sets of features may be better represented in different geometric spaces.
Cartesian products of manifolds enable the simultaneous processing of data representations across multiple manifolds, offering several unique perspectives on the dataset at once. 
Park et. al \cite{sangeonpark_2023_neural} explored the use of hyperbolic geometries in the final layers of models for jet analysis, displaying the hierarchies of embedding spaces. 
We expand this approach by developing model architectures compatible with jets represented in \(\mathcal P \mathcal M\)s throughout the model. 

We define two representations of jets in machine learning models: the \textit{particle-level} and \textit{jet-level} representations. In the particle-level representation, a jet is treated as a collection of individual particles, each embedded within a manifold. By contrast, the jet-level representation condenses this information into a single latent vector that represents the entire jet after aggregating the particle-level data. 
In this work, we explore the use of \(\mathcal P \mathcal M\) representations for both the particle-level and jet-level representations.
To utilize \(\mathcal P \mathcal M\) representations, we adapt multilayer perceptron (MLP) and transformer models to process data with \(\mathcal P \mathcal M\) representations at the particle-level and jet-level. 
We develop highly generalized architectures that allow seamless use of any \(\mathcal P \mathcal M\) representation and systematic searches for the optimal one.
Our results on jet tagging demonstrate that \(\mathcal P \mathcal M\) representations generally achieve comparable or superior performance, with the most pronounced improvements observed for highly hierarchical jets.

This work underscores the importance of tailoring representations to the data's intrinsic structure. 
Our findings in the context of jet tagging in particle physics provide a first promising example of the broader potential of $\mathcal P \mathcal M$ machine learning in unraveling the complexities of natural data.

\paragraph{Paper Outline}The paper is organized as follows. In~\Cref{sec:math_prereq} we provide a summary of the mathematical perquisites necessary to develop this work; in~\Cref{sec:manifold_ml} we present rigorous non-Euclidean analogs for machine learning operations; in~\Cref{sec:model_arch} we present the \(\mathcal P \mathcal M\)-MLP and \(\mathcal P \mathcal M\)-Transformer models. We perform numerical experiments on particle jets data from collider experiments as use case. Exploratory tests for both architectures are presented in~\Cref{sec:perf_compare}; results for benchmarking the $\mathcal P \mathcal M$ transformer on the typical particle physics tasks of top tagging, are reported in~\Cref{sec:bench}. Concluding remarks and outlook are given in~\Cref{sec:conclusions}

\section{Mathematical Prerequisites}
\label{sec:math_prereq}
\begin{wrapfigure}{r}{0.65\textwidth}
    \centering
    \vspace{-0.7cm}
    \includegraphics[width=0.7\linewidth]{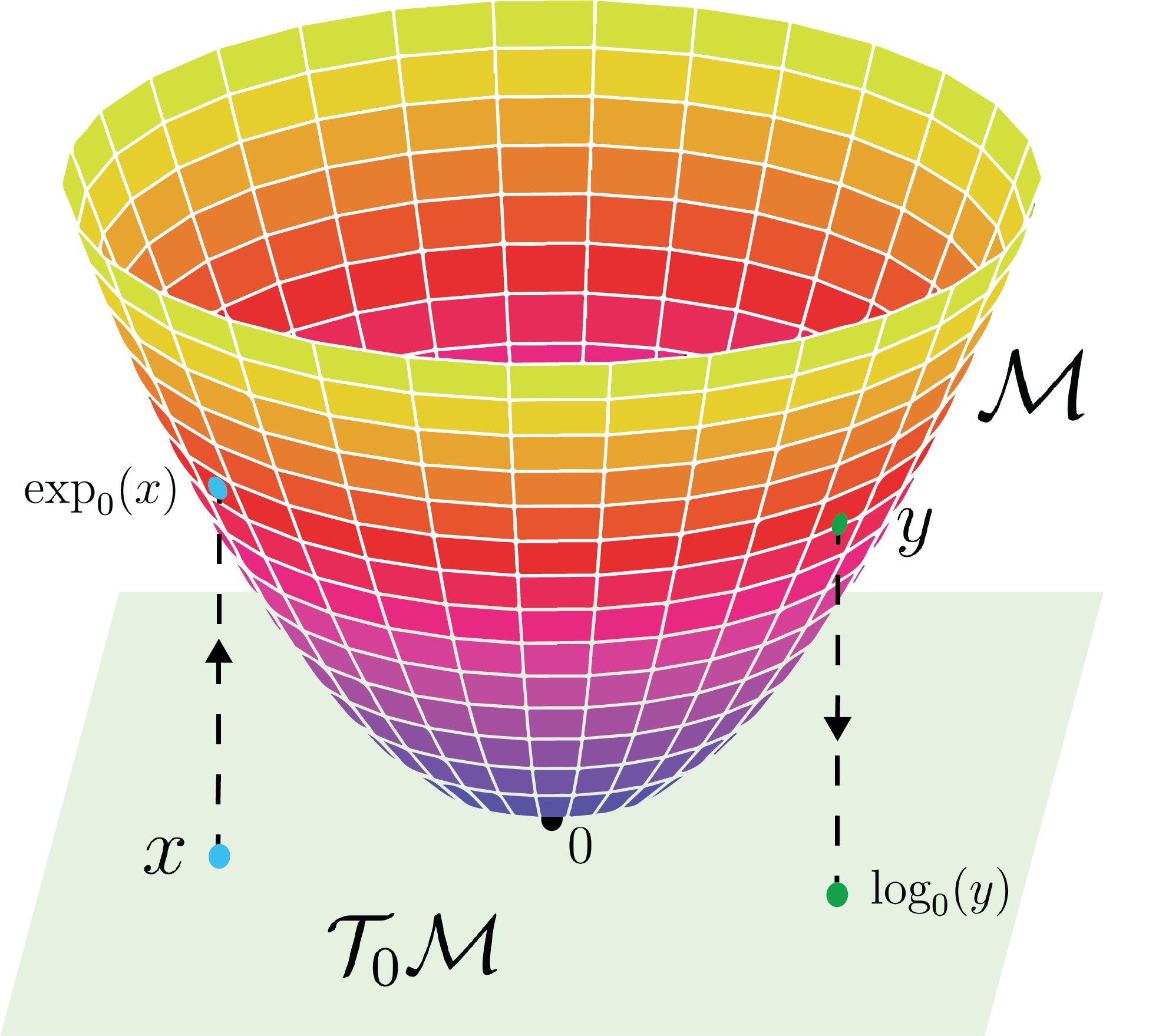}
    
    \caption{A graphical depiction of a manifold $\mathcal M$ together with the tangent space at the point $x = 0$, denoted $\mathcal T_0 \mathcal M$ and shown in green. We further illustrate the use of the exponential map, which for $x \in \mathcal T_0 \mathcal M$ takes $x \rightarrow \text{exp}_0(x) \in \mathcal M$, and the logarithmic map, which for $y \in  \mathcal M$ takes $y \rightarrow \text{log}_0(x) \in \mathcal T_0 \mathcal M$. A portion of this image is adapted from \cite{wiki:parabola}.}
    \label{fig:tan_space}
    \vspace{-2em}
\end{wrapfigure}

In this section we provide an overview of the main concepts required for the formulation of machine learning on non-Euclidean geometries. The reader is referred to \cite{oneill_1997_elementary,ungar_2005_analytic} for further details.
\paragraph{Riemannian Manifolds}
A \(d\)-dimensional Riemannian manifold \((\mathcal{M}, g)\) is a smooth manifold, denoted as \(\mathcal M^d\), together with a Riemannian metric \(g\) which determines the curvature \( \kappa \) at each point \( x \in \mathcal{M}^d \).
In this work, we focus on constant curvature manifolds, where curvature is uniform across the entire space: \(\kappa < 0\) for hyperbolic spaces \(\mathbb H\), \(\kappa = 0\) for Euclidean spaces \(\mathbb R\), and \(\kappa > 0\) for spherical spaces \(\mathbb S\).
Since \(\mathcal M^d\) is a smooth manifold, each point \(x \in \mathcal M^d\) is equipped with a tangent space \(\mathcal T_{x}\mathcal M^d \subseteq \mathbb{R}^d\), providing a local, linear approximation of the manifold near \(x\). The exponential (\cref{eq:exp_u}) and logarithmic (\cref{eq:log_u}) maps are employed throughout the \(\mathcal P \mathcal M\) machine learning models presented to map latent vectors between the Euclidean space and the manifold in use, as illustrated in \Cref{fig:tan_space}. 

\begin{equation}
\label{eq:exp_u}
\exp_u(\cdot): \mathcal{T}_{u}\mathcal{M}^d \rightarrow \mathcal{M}^d, \quad
\exp_u(v) = \gamma_{u, v}(1) = u \oplus_\kappa \tan_\kappa\left(\frac{\|v\|_u}{2}\right) \frac{v}{\|v\|_2}
\end{equation}
\vspace{-0.25em}
% Logarithmic map centered at u
\begin{equation}
\label{eq:log_u}
\log_u(\cdot): \mathcal{M}^d \rightarrow \mathcal{T}_{u}\mathcal{M}^d, \quad
\log_u(w) = \frac{2}{\lambda_u^\kappa}
        \tan_\kappa^{-1}(\|(-u) \oplus_\kappa w\|_2)
        \frac{(-u) \oplus_\kappa w}{\|(-u) \oplus_\kappa w\|_2}
\end{equation}

\paragraph{Product Spaces}
As outlined in \Cref{sec:intro}, we employ a general data representation formed by combining several distinct manifolds with curvatures. The product space \( \mathcal{P} \) is defined by linking these manifolds using the Cartesian product:
\begin{equation}
    \mathcal P = \mathcal M_{\kappa_1}^{d_1} \times \mathcal M_{\kappa_2}^{d_2} \times ... \times \mathcal M_{\kappa_n}^{d_n}
\end{equation}
The total dimension of \(\mathcal P\) is equal to the sum of each individual manifold, which we denote as  \(d = \sum_{i=1}^n d_i\). We refer to product space representations as \textit{product manifolds} (\(\mathcal P \mathcal M\)).

\begin{figure}[h]
    \centering
    % First subfigure
    \begin{subfigure}[t]{0.45\textwidth}  % [t] aligns subfigures at the top
        \centering
        \includegraphics[width=0.75\linewidth]{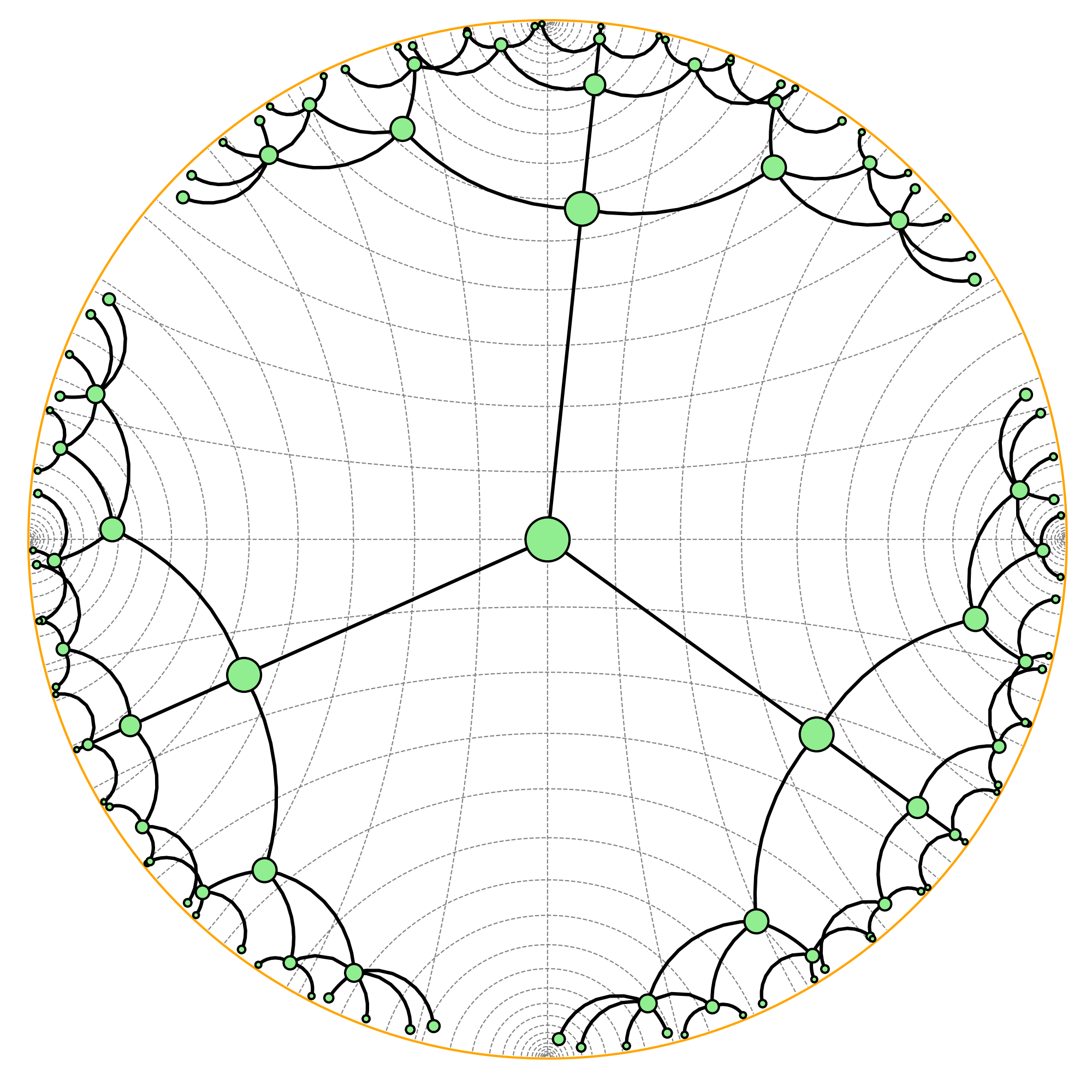}
        \caption{A rooted tree with a branching ratio of 3 and a depth of 4 is plotted in the 2D Poincaré Disk, illustrating the exponential growth of area and volume characteristic of hyperbolic spaces.}
        \label{fig:poincare_ball}
    \end{subfigure}
    \hfill % Add some horizontal spacing
    % Second subfigure
    \begin{subfigure}[t]{0.45\textwidth}  % [t] aligns subfigures at the top
        \centering
        \includegraphics[width=\linewidth]{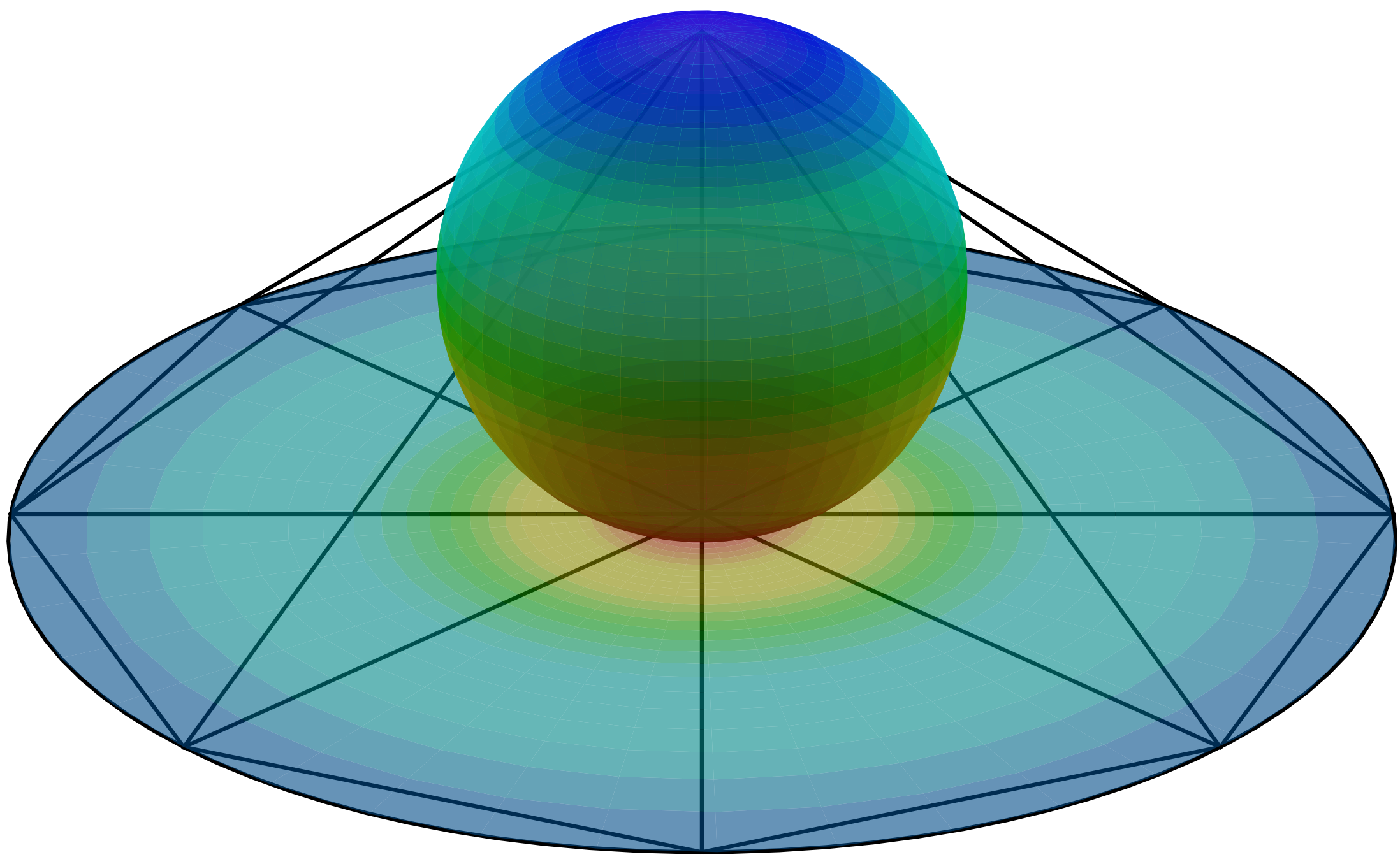}
        \caption{Stereographic projection of 2D sphere onto 2D plane with coloring according to stereographic mapping.}
        \label{fig:sphere_proj}
    \end{subfigure}
    \caption{Illustrations of non-Euclidean manifold representations used in this work}
\end{figure}

\paragraph{Representation Models} 
We utilize stereographic projection representations of 
hyperbolic and spherical spaces, the Poincaré ball model for hyperbolic spaces, shown in \Cref{fig:poincare_ball}, and the stereographic spherical projection model for spherical spaces, shown in \Cref{fig:sphere_proj}. We see these as the natural choice for \(\mathcal P \mathcal M\) approaches as they have a unified gyrovector formalism developed by Bachmann et. al \cite{bachmann_2020_constant}.

\paragraph{Gyrovector Spaces}
\label{sec:gyro}

\begin{table}[t]
\centering
\begin{tabular}{|l|c|c|}
\hline
Euclidean Operations & Gyrovector Analogs & G./E. FLOPS Ratio\\
\hline
\rule[-1.5ex]{0pt}{4.5ex}\(x + y\) & \(x \oplus_\kappa y =
        \frac{
            (1 - 2 \kappa \langle x, y\rangle - \kappa \|y\|^2_2) x +
            (1 + \kappa \|x\|_2^2) y
        }{
            1 - 2 \kappa \langle x, y\rangle + \kappa^2 \|x\|^2_2 \|y\|^2_2
        }\) & 5.1
\\
\hline
\(\rule[-1.5ex]{0pt}{4ex} r \mathbf{x}\) & \(r \otimes_\kappa x
        =
        \tan_\kappa(r\tan_\kappa^{-1}(\|x\|_2))\frac{x}{\|x\|_2}\) & 3
        \\
\hline
\(\rule[-1.5ex]{0pt}{4ex}|x-y|\) & \(d_\kappa(x, y) = 2\tan_\kappa^{-1}(\|(-x)\oplus_\kappa y\|_2)\) & 4.5\\
\hline
\(Mx\) & \(M \otimes_\kappa x = \tan_\kappa\left(
            \frac{\|Mx\|_2}{\|x\|_2}\tan_\kappa^{-1}(\|x\|_2)
        \right)\frac{Mx}{\|Mx\|_2}\) & 2.8 \\
\hline
\end{tabular}
\vspace{0.1in}
\caption{We show Euclidean vector operations and their gyrovector analogs. Floating-point operations per second (FLOPS) ratios are presented for 4D vector/gyrovector operations, bringing an \(\mathcal O(n)\) FLOP correction.}
\label{tab:mobius_ops}
\end{table}

Gyrovector spaces were developed by Ungar for hyperbolic spaces (\(\kappa < 0\)) \cite{ungar_2005_analytic} and extended to stereographic spherical projections geometries (\(\kappa > 0 \)) by Bachmann et. al \cite{bachmann_2020_constant}. 
When performing calculations in non-Euclidean spaces in this work, Euclidean vector operations are replaced by gryovector operations, shown in \Cref{tab:mobius_ops} which highlights the increased complexity required for gyrovector operations. 

\paragraph{Gromov-\(\delta\) Definition of Curvature}
\label{sec:gromov}
Gromov-\(\delta\) hyperbolicity estimates the degree of hierarchical structure in the dataset \cite{khrulkov_hyperbolic}, where the hyperbolicity \(\delta\) is chosen such that for any points \(\mathbf{a}, \mathbf{b}, \mathbf{c}, \mathbf{d}\) in a \(\delta\)-hyperbolic, they satisfy,
\begin{equation}
\label{eq:gromov}
    (\mathbf a,\mathbf b)_{\mathbf d} \geq \text{min}\{(\mathbf b,\mathbf c)_{\mathbf d}, (\mathbf a,\mathbf c)_{\mathbf d} \} - \delta, \quad \forall \mathbf a, \mathbf b, \mathbf c, \mathbf d \in \mathcal{M}
\end{equation}
\begin{equation}
\label{eq:gromovprod}
    (\mathbf x,\mathbf y)_{\mathbf z} = \frac 1 2 (d(x,z) + d(y,z) - d_{\mathbb R}(x,y))
\end{equation}
where \((\mathbf x,\mathbf y)_{\mathbf z}\) is the Gromov product in \cref{eq:gromovprod} for the distance metric \(d(\cdot, \cdot)\). 
The Gromov\(-\delta\) hyperbolicity is related to the scalar curvature of the manifold through an inverse square relation. 

\paragraph{Einstein Midpoint}
\label{sec:einstein_mid}
Data aggregation operations are crucial to many modern machine learning architectures but non-trivial for manifold data. Following \cite{aglarglehre_2018_hyperbolic}, we utilize the weighted Einstein midpoint to aggregate vectors on manifolds. For vectors $x_1,x_2, ...,x_n \in \mathcal M^d_\kappa$ and weights $\alpha_1,\alpha_2, ...,\alpha_n \in \mathbb R$ the Einstein midpoint is defined in \cref{eq:einsteinmidpoint}.

\begin{equation}
\label{eq:einsteinmidpoint}
    m(\mathbf x_1,\mathbf x_2,..., \mathbf x_n; \alpha_1,\alpha_2, ...,\alpha_n) = \frac 1 2 \otimes_\kappa \sum_i^n \frac{\lambda_\kappa(\mathbf x_i) \alpha_i}{\sum_j^n (1-\lambda_\kappa (\mathbf x_j)) \alpha_j}\mathbf x_i
\end{equation}
where $\lambda_\kappa(\mathbf x_i)$ is the conformal factor \cite{ungar_2005_analytic} for the point $x_i \in \mathcal M^d_\kappa$ defined as

\begin{equation}
    \label{eq:conformal factor}
    \lambda_\kappa(\mathbf x_i) = \frac {1}{\sqrt{1 - \kappa^2 \| \mathbf x_i\|^2}}
\end{equation}

\section{Manifold Machine Learning}
\label{sec:manifold_ml}
In this section, we summarize the modifications to traditional machine learning operations necessary to build mathematically rigorous models across all constant curvature Riemannian manifolds. 
Our implementation is built using \textsc{Geoopt} \cite{geoopt2020kochurov}, an open source package for Riemannian optimization in \textsc{PyTorch}.

\paragraph{Fully Connected Layers}
\label{sec:man_FC}
Ganea et. al \cite{octavianeugenganea_2018_hyperbolic} proposed utilizing gyrovector operations to develop fully connected layers applicable for Riemannian manifolds of all curvature. For weights \(W \in \mathbb{R}^{m,n}\), bias \(\mathbf b \in \mathcal M_\kappa^m\), and input \(\mathbf x \in \mathcal M_\kappa^n\) we can calculate the output \(\mathbf y \in \mathcal M_\kappa^m\) as: 
\begin{equation}
    FC(\mathbf x; n, m, \kappa) = W \otimes_\kappa \mathbf x \oplus_\kappa \mathbf b
\end{equation}
where \(\otimes_\kappa\) and \(\oplus_\kappa\) are the gyrovector matrix multiplication and vector addition, respectively. We will refer to a stack of manifold fully connected layers in \(\mathcal P \mathcal M\) models as \(\mathcal M\)-MLP (manifold MLP) and we reserve MLP for the Euclidean MLP.

\paragraph{Activation Functions}
There are many approaches in the literature to formulating activation functions in non-Euclidean spaces \cite{cohen2018sphericalcnns,NEURIPS2019_103303dd,peng2021hyperbolicdeepneuralnetworks,bachmann_2020_constant}. In our testing we have found that \textsc{ReLU} activation functions yields minor improvements compared to no activation functions. This result is unsurprising as activations inject non-linearity, however, non-Euclidean spaces are already non-linear. Furthermore, we found that activation functions in the tangent space degrade performance.

\paragraph{LayerNorm}
Implementing normalization layers for non-Euclidean machine learning has remained a challenge \cite{peng2021hyperbolicdeepneuralnetworks}. Rigorous methods such as those proposed by \cite{lou2021differentiatingfrechetmean} rely on iterative calculations which are a detrimental bottleneck for deep models. To avoid these challenges, we implement \textsc{LayerNorm}
in the tangent space. This provides a simple method for normalization across all manifolds considered without significant negative impacts on training or inference times. 

\begin{figure}[h]
    \centering
    % First figure in the left
    \begin{minipage}[t]{0.39\linewidth}
        \centering
        \includegraphics[width=\linewidth]{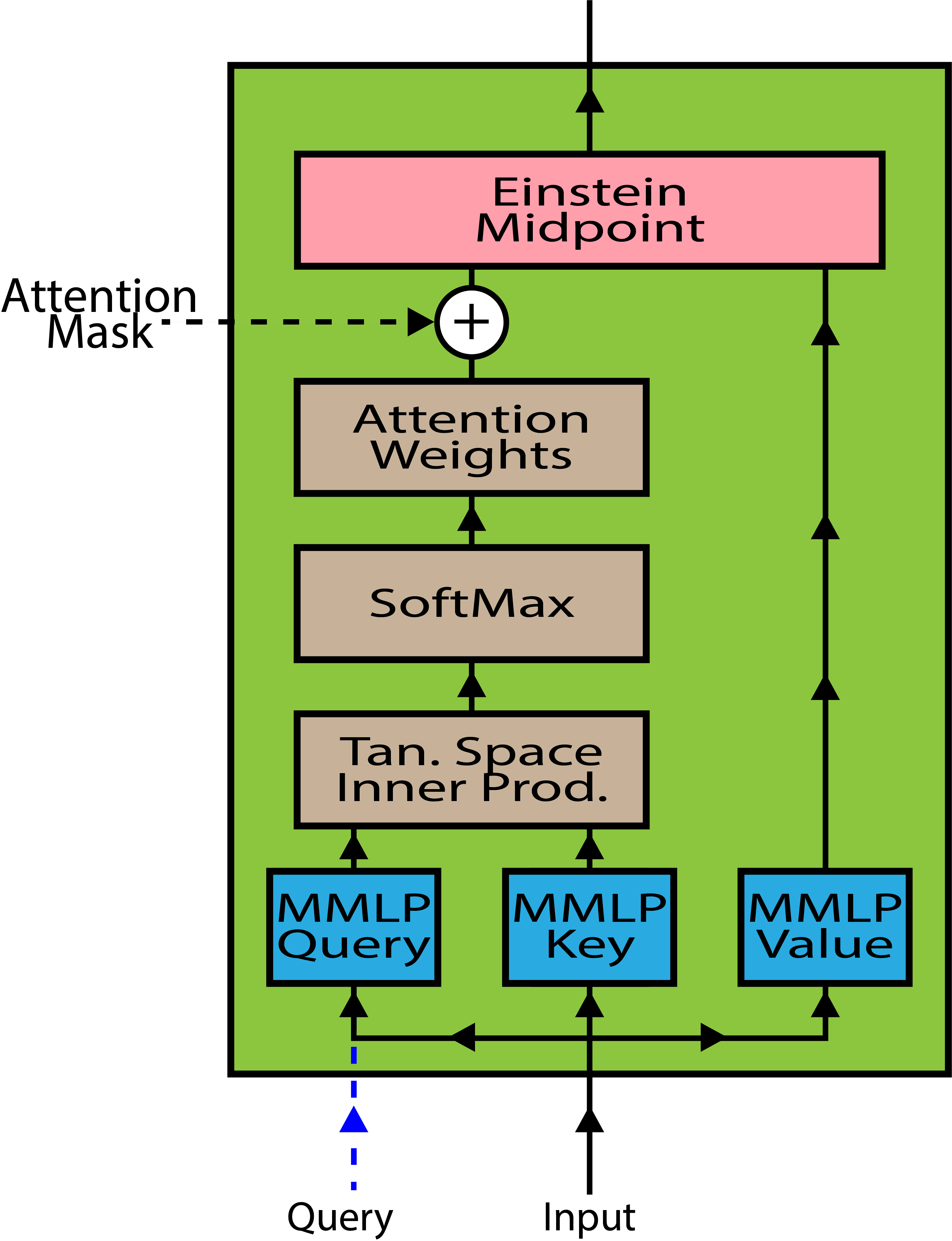}
        \caption{An illustration of the manifold attention layer. We use MMLPs to determine query, key, and value vectors. Attention weights are calculated from tangent space inner product of query and key vectors and attention probabilities are determined through a \textsc{SoftMax} operation. We apply an attention mask if supplied to attention probabilities and aggregate using the Einstein midpoint weighted by the attention probabilities.}
        \label{fig:man_att}
    \end{minipage}
    \hfill % Adds some space between the two figures
    % Second figure on the right
    \begin{minipage}[t]{0.59\linewidth}
        \centering
        \includegraphics[width=\linewidth]{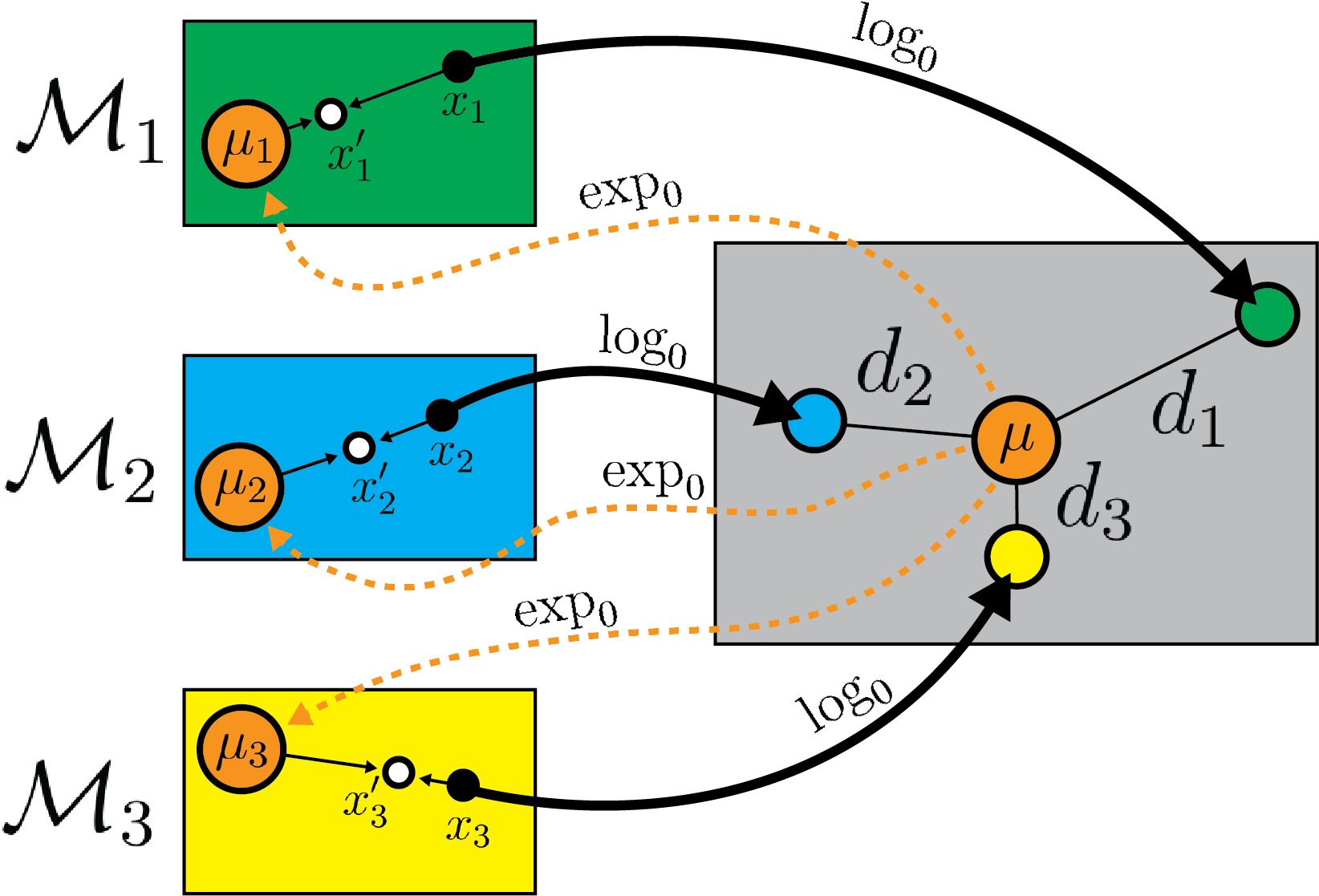}
        \caption{An illustration of the inter-manifold attention calculation for a \(\mathcal P \mathcal M\) data point \(x = (x_1, x_2, x_3) \in \mathcal P = \mathcal M_1 \times \mathcal M_2 \times \mathcal M_3\). The three manifolds \(\mathcal M_1,\mathcal M_2,\mathcal M_3\) are color coded as green, blue, and yellow, respectively, and the weighted midpoint \(\mu\) and projected means \(\mu_1, \mu_2,\mu_3\) are shown in orange. Logarithmic maps are shown as solid black lines and exponential maps are shown as dotted orange lines. The updated \(\mathcal P \mathcal M\) data point \(x' = (x_1', x_2', x_3')\) is shown in white in each manifold.}
        \label{fig:inter_man_att}
    \end{minipage}
\end{figure}

\paragraph{Attention Mechanism}
\label{sec:man_att}
For multi-headed attention, we perform all reshape operations in the tangent space \(\mathcal T_0 \mathcal M\), projecting the reshaped result back onto the manifold \(\mathcal M\). 
We calculate key ($K$), query ($Q$), and value ($V$) for each attention head using $\mathcal M$-MLPs, defined in \cref{sec:man_FC}. 
Attention weights are calculated using the tangent space inner product operation shown in \cref{eq:man_att_inner}. We compare this tangent space attention mechanism to the distance-based attention mechanism of \cite{aglarglehre_2018_hyperbolic} in \Cref{sec:perf_compare}. 
We calculate the attention probabilities using a \textsc{SoftMax} activation function and aggregate the value vectors (V) using the Einstein midpoint weighted by the attention probabilities, defined in \Cref{sec:einstein_mid}.
An illustration of the full attention calculation is provided in \Cref{fig:man_att}.

\begin{equation}
\label{eq:man_att_inner}
    a_{ij} = \langle \log_0(q_i), \log_0(k_j) \rangle 
\end{equation}

\paragraph{Inter-Manifold Attention}
\label{sec:inter_man}
Sun et al. \cite{sun_2022_a} formulated inter-manifold attention to quantify relationships between data represented across distinct manifolds within the product space. 
Their approach maps all manifolds representations in the \(\mathcal P \mathcal M\) to a tangent space of equal dimension and calculates inter-manifold attention probabilities as the traditional Euclidean attention between these tangent space representations. 
Using these weights, they scale the product space representations in each manifold by the corresponding attention probabilities using gyrovector scalar multiplication. 
In this approach, information transfer between manifolds relies solely on scaling each representation by attention probabilities. 
We find that in complex models, this rescaling operation is too simple to bring performance gains. 
We overcome these challenges by developing a novel method of inter-manifold attention illustrated in \Cref{fig:inter_man_att}.

For a point \(x \in \mathcal P\) with \(\mathcal P = \mathcal M_1 \times \mathcal M_2 \times ... \times \mathcal M_n\) we map all representations to their respective tangent spaces and we enforce all tangent spaces to be equal dimension ($\rm k$) through FC layers. 
This yields a set of points in a Euclidean space \(x_i^{\mathcal T} \in \mathbb R^k\) for \(i \in [1,...,n]\). 
Then we calculate the weighted midpoint \(\mu\) for all \(x_i^{\mathcal T}\) where the weights are determined through a small MLP taking as input the concatenated tangent space representations (\(x_i^{\mathcal T}\)) and outputting a \(n\)-dimensional weight vector. We calculate the normalized distances from each (\(x_i^{\mathcal T}\)) to the weighted midpoint as \(d_i = \textsc{SoftMax}\|x_i^{\mathcal T}-\mu\|\). 
To update an individual manifold representation \( x_i \in \mathcal M_i \), we first project $\mu$ onto the manifold through the exponential map and denote this projected mean as \(\mu_i\). 
We updated \(x_i\) through the weighted midpoint of \(x_i\) and \(\mu\) with weights \(1, d_i\), respectively.

This new method perturbs a point in the manifold \(x_i\) in the direction of the projected learned midpoint \(\mu_i\) with a strength dependent on the relative distance \(d_i\) in the tangent space. 
If a point has a large distance to the midpoint, the perturbation is strong, while if a point is close to the midpoint, the perturbation is weak. We demonstrate a direct comparison of our novel method for inter-manifold attention compared to that of Sun et al. in \Cref{sec:perf_compare}.

\section{Model Architectures}
\label{sec:model_arch}
In this section we describe the \(\mathcal P\mathcal M\)-MLP and \(\mathcal P\mathcal M\)-Transformer architectures used in the numerical experiments on particle jets presented in Section~\ref{sec:perf_compare} and~\ref{sec:bench}. 
\subsection{Product Manifold Multilayer Perceptron (\(\mathcal P \mathcal M\)-MLP)}
\label{sec:PMNN}
\begin{figure}[h]
    \centering
    \includegraphics[width = \linewidth]{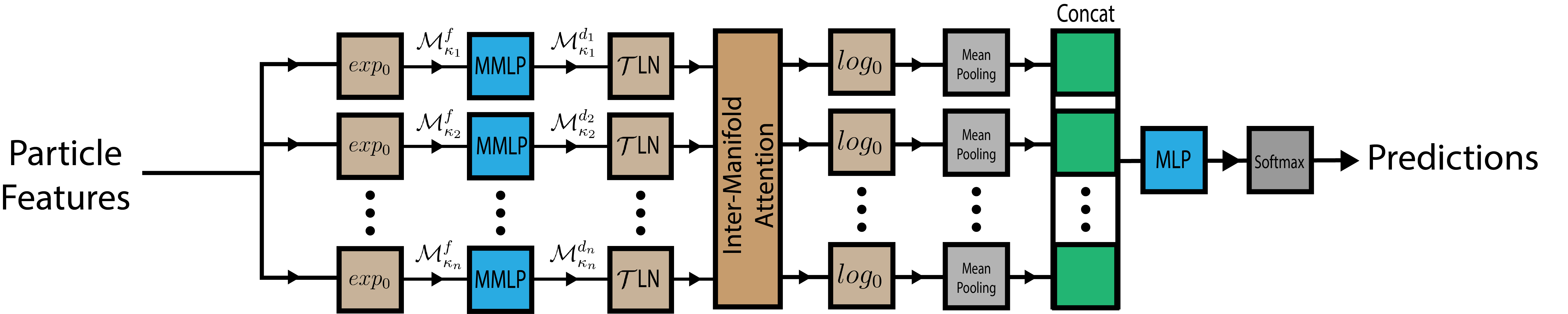}
    \caption{The schematic of the \(\mathcal P \mathcal M\)-MLP model shows jet constituents mapped to the particle-level \(\mathcal P \mathcal M\) \( \mathcal{P} \), processed through $\mathcal M$-MLP with ReLU activations and an inter-manifold attention layer. The outputs are mean aggregated in the tangent space and concatenated, forming a jet-level latent vector, which is processed in Euclidean space with a MLP for predictions.}
    \label{fig:mlp_diagram}
\end{figure}

The $\mathcal P \mathcal M$-MLP architecture is a simple $\mathcal M$-MLP based model architecture, aiming to highlight the impact of the chosen representation. 
The architecture is configurable for any possible product-manifold at the particle-level. We restrict the jet-level representations to be Euclidean to simplify the model architecture. 
A schematic view of the model's architecture is shown in \Cref{fig:mlp_diagram}. 

\vspace{-1em}
For an input of jet constituents, each with $f$ particle-level features, we first map onto the particle-level \(\mathcal P \mathcal M\) ($\mathcal P$) using the exponential map $\text{exp}_0(\cdot)$. 
For each representation $\mathcal M_{\kappa_i}^{d_i} \in \mathcal P$, we apply a three-layer $\mathcal M$-MLP with \textsc{ReLU} activations which embeds the particle features in $d_i$ dimensions. If the representation consists of two or more manifolds, we apply an inter-manifold attention layer.\footnote{For the \(\mathcal P \mathcal M\)-MLP we use the scaling-based inter-manifold attention from Sun et al. \cite{sun_2022_a} to explore the simplest possible architecture. Our proposed method of inter-manifold attention employed a learned weighted midpoint, which increases total parameters significantly for small models.} 
To aggregate the particle-level features to a jet-level representation, we map each \(\mathcal P \mathcal M\) representation $\mathcal M^{d_i}$ to the tangent space $\mathcal T_0\mathcal M^{d_i}$ using the logarithmic map $\text{log}_0(\cdot)$. 
Now in Euclidean space, we aggregate particle-level representations using mean aggregation. To form the jet-level representation we concatenate aggregated representations from each manifolds in $\mathcal P$. 
We pass this jet representation into a final two-layer Euclidean MLP followed by a \textsc{SoftMax} activation function for jet classification. When benchmarking \(\mathcal P \mathcal M\)-MLP models in \Cref{sec:perf_compare} we will refer to different \(\mathcal P \mathcal M\)-MLP models by their internal representation $\mathcal P = \mathcal M^{d_1} \times ...  \times \mathcal  M^{d_n}$. 

\subsection{Product Manifold Transformer ($\mathcal P \mathcal M$-Transformer)}
\label{sec:PMTrans}

\begin{figure}[h]
    \centering
    \includegraphics[width=1\textwidth]{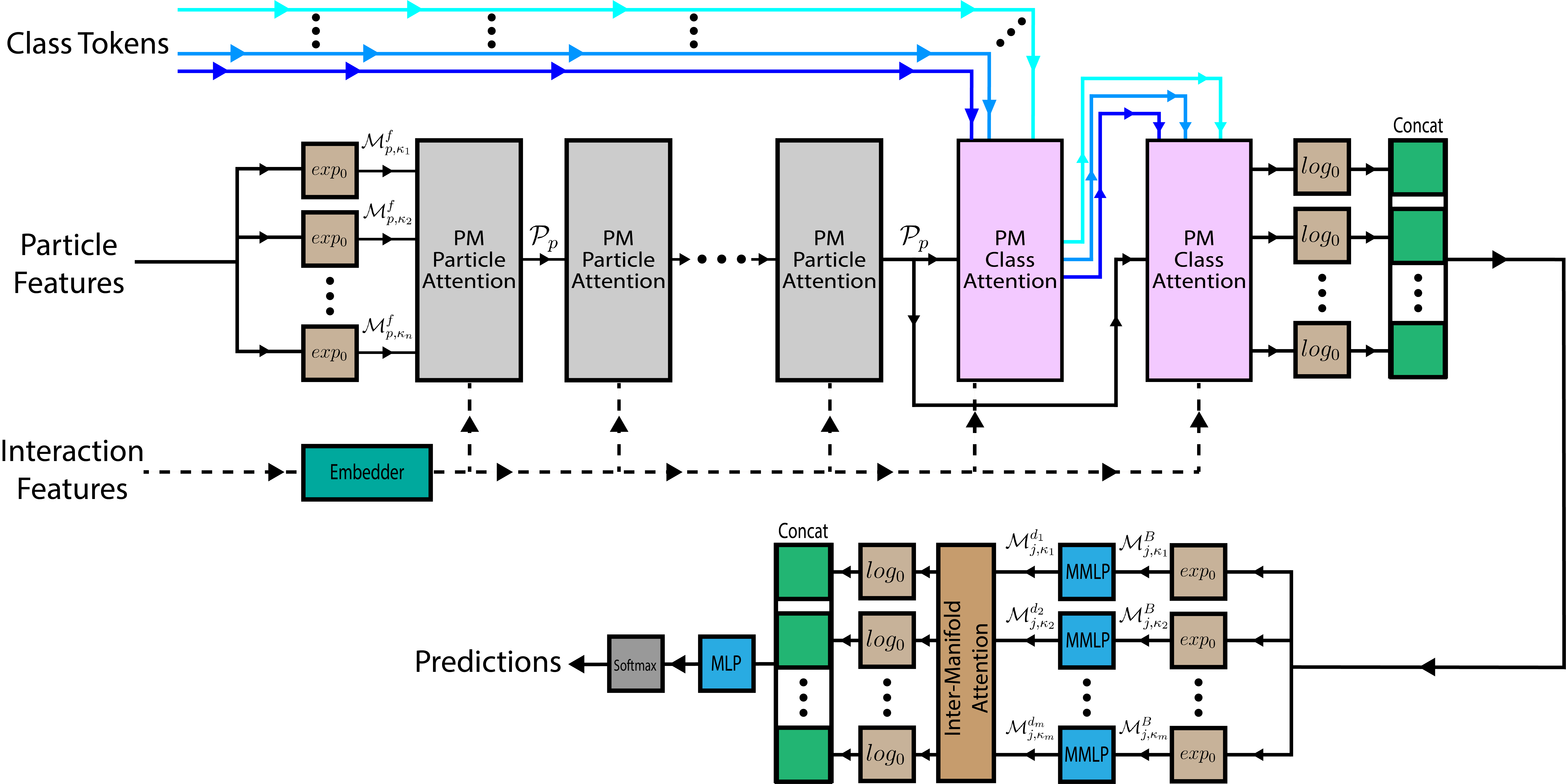} % Adjust width to 90% of the text width
    \caption{The schematic of the \( \mathcal{P}\mathcal{M} \)-Transformer model shows jet constituents mapped to the particle-level \(\mathcal P \mathcal M\) \( \mathcal{P}_p \), processed through \( \mathcal{P}\mathcal{M} \) particle and class attention blocks. The aggregated outputs are concatenated in the tangent space, forming a jet-level latent vector, which is mapped to the jet-level \(\mathcal P \mathcal M\) \( \mathcal{P}_j \) and processed with \( \mathcal{M} \)-MLP and inter-manifold attention layers. The final jet-level latent vectors are concatenated in tangent space and passed through an MLP for predictions.}
    \label{fig:trans}
\end{figure}
For the $\mathcal P \mathcal M$-Transformer we utilize a generalized model architecture shown in \Cref{fig:trans} which is inspired by the particle transformer (ParT) \cite{qu_2022_particle}. $\mathcal P \mathcal M$ models are designed to be compatible with any possible particle-level and the jet-level representation. For clarity, we differentiate between the particle-level and jet-level representation through
\begin{equation}
    \begin{aligned}
    \text{Particle-Level Representation: } \mathcal P_p = \mathcal M_{p,\kappa_1}^{d_1} \times \mathcal M_{p,\kappa_2}^{d_2} \times ... \times \mathcal M_{p,\kappa_n}^{d_n}\\
    \text{Jet-Level Representation: } \mathcal P_j = \mathcal M_{j,\kappa_1}^{d_1} \times \mathcal M_{j,\kappa_2}^{d_2} \times ... \times \mathcal M_{j,\kappa_m}^{d_m}
    \end{aligned}
\end{equation}

\subsubsection{Particle-Level Processing}
The particle-level data is mapped onto each manifold and processed with a sequence of $\mathcal P \mathcal M$ particle attention blocks and $\mathcal P \mathcal M$ class attention blocks. 

\begin{wrapfigure}{r}{0.65\textwidth}
\captionsetup{type=figure}
\centering
\includegraphics[width=\linewidth]{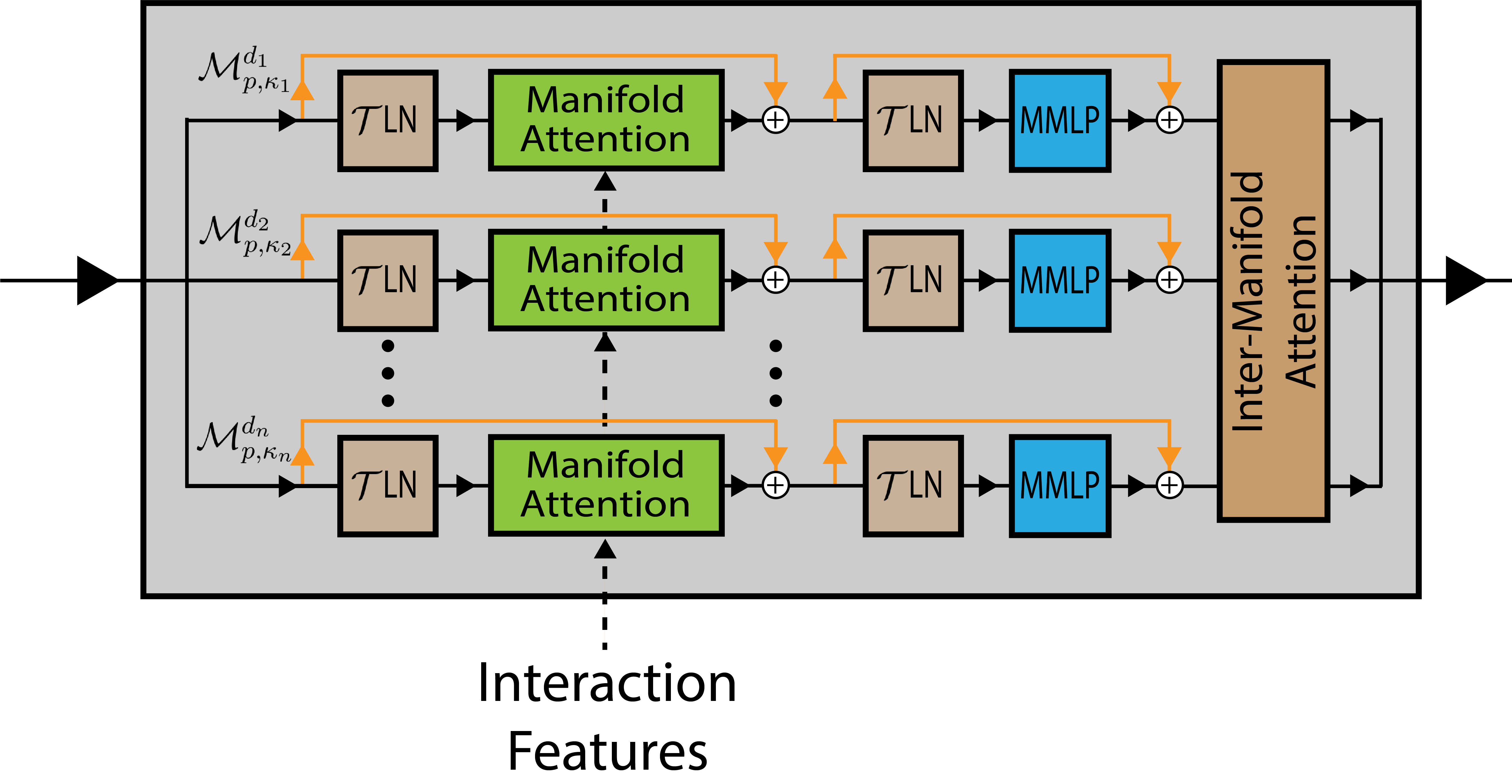}
\caption{Product Manifold ($\mathcal P \mathcal M$) Particle Attention block.\label{fig:part_att}}
\vspace{-1em}  % Adjust spacing below the caption as needed
\end{wrapfigure}

\textbf{$\mathcal P \mathcal M$ particle attention block: }A $\mathcal P \mathcal M$ particle attention block integrates a traditional attention mechanism within each representation. 
A schematic representation of the block is given in \Cref{fig:part_att}. 
The \(\mathcal P \mathcal M\) particle block takes as input the particle-level \(\mathcal P \mathcal M\) latent vector \(x \in \mathcal{P}_p\) and applies a tangent space \textsc{LayerNorm} (\(\mathcal{T}\)LN), a manifold attention layer, and a $\mathcal M$-MLP layer in sequence to each representation.
The $\mathcal M$-MLP consists of two fully connected layers with \textsc{ReLU} activations and dropout. 
Residual connections, depicted as orange arrows in \Cref{fig:part_att}, are implemented through the gyrovector addition.

As a final output, we compare representations across the $\mathcal P \mathcal M$ with an inter-manifold attention layer (\Cref{sec:inter_man}). In the sequence of \(\mathcal{P} \mathcal{M}\) particle attention blocks, inter-manifold attention is included by default on every even-numbered encoder, except for the first and last ones.
To enhance numerical stability, we project latent vectors onto the manifold following the attention calculations, Möbius addition for residual connections, and both $\mathcal{T}$LN layers.

For particle jets, physically meaningful relationships between jet constituents can be derived from their 4-momentum which can guide the attention calculation through the attention mask. 
This is achieved by applying an attention mask based on pairwise features of the jet constituents, referred to by Qu et. al \cite{qu_2022_particle} as \textit{interaction features}. The motivation is to enforce stronger attention relationships between jet constituents that are correlated due to the underlying physical processes forming the jet.
Starting from the 4-momentum $(E, \vec p)$ of two particles, $p_1$ and $p_2$, we define the interaction features as
\begin{equation}
\begin{aligned}
\Delta &= \sqrt{(\eta_{1} - \eta_{2})^2 + (\phi_{1} - \phi_{2})^2}, \\
k_{T} &= \min(p_{T,1}, p_{T,2}) \Delta, \\
z &= \min(p_{T,1}, p_{T,2}) / (p_{T,1} + p_{T,2}), \\
m^2 &= (E_{1} + E_{2})^2 - \| \vec p_{1} + \vec p_{2} \|^2,
\end{aligned}
\end{equation}
where $\eta$ is the rapidity, $\phi$ is the azimuthal angle, and $\|\cdot\|$ is the Euclidean norm. To normalize the interaction features, we apply a logarithm resulting in $(\ln(\Delta), \ln(k_T), \ln(z), \ln(m^2))$. These features are further processed using a 4-layer pointwise 1D convolution with $(64,64,64,h)$ channels, where $h$ is the number of attention heads, with GELU activations and batch normalization between each convolutional layer \cite{qu_2022_particle}.

% Second figure: Class Attention Block
\begin{wrapfigure}{r}{0.65\textwidth}
\vspace{-0.8cm}
\captionsetup{type=figure}
\centering
\includegraphics[width=\linewidth]{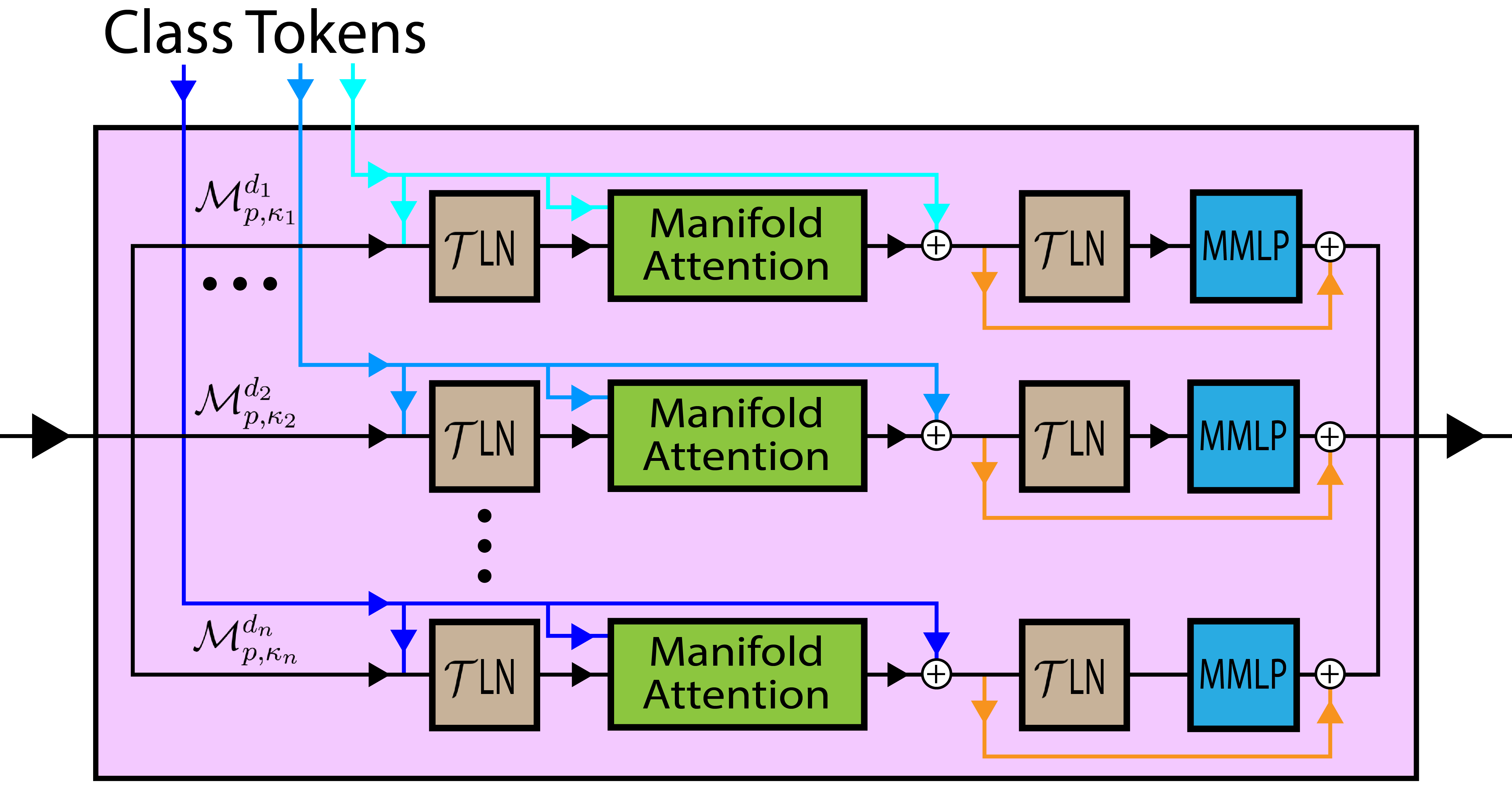}
\caption{Product Manifold ($\mathcal P \mathcal M$) Class Attention block\label{fig:jet_att}}
\end{wrapfigure}

\textbf{$\mathcal P \mathcal M$ class attention block: }Following the $\mathcal P \mathcal M$ particle attention blocks, we apply a sequence of $\mathcal P \mathcal M$ class attention blocks. $\mathcal P \mathcal M$ class attention blocks follow the architecture of the $\mathcal P \mathcal M$ particle attention block but uses a global class token for each manifold in the particle-level representation. For an input $x \in \mathcal P_p$ to a $\mathcal P \mathcal M$ class attention blocks, the traditional manifold attention is modified by replacing query, key, and value with
\\
\begin{equation}
\begin{aligned}
    Q_{\mathcal M_i} &= W_{q,i} \otimes \mathbf \mathbf x_{\text{class},i} \oplus \mathbf{b}_{q,i} \\
    K_{\mathcal M_i} &= W_{k,i} \otimes \mathbf z_{i} \oplus \mathbf b_{k,i} \\
    V_{\mathcal M_i} &= W_{v,i} \otimes \mathbf z_{i} \oplus \mathbf{b}_{v,i}
\end{aligned}
\end{equation}

where $x_{\text{class},i} \in \mathcal M^{d_i}$ is the class token, $z_i = [x_{\text{class},i},x_{i}]$ is the input concatenated with the class token and $W_{\{q,k,v\},i}$ and $b_{\{q,k,v\},i}$ are $\mathcal M$-MLP weights and biases, respectively. 
This modified structure aggregates particle latent vectors through the attention comparison of $z_i$ with the global token $x_{\text{class},i}$. 
This approach follows the CaiT architecture from \cite{touvron2021goingdeeperimagetransformers}.  
To conclude the particle-level processing we map all representations to their corresponding tangent spaces using the $log_0(\cdot)$ map and concatenate each respective $x_i$, forming a jet-level latent vector in $\mathbb R^B$, for $B = \sum_{i=1}^n d_i$.

\subsubsection{Jet-Level Processing}
The particle-level processing described above is followed by a jet-level processing stage, mapping the aggregated particle-level data onto the jet-level representation $\mathcal P_j$. We apply a $\mathcal M$-MLPs for each manifold in the jet-level representation and an inter-manifold attention layer (\Cref{sec:inter_man}) if $\mathcal{P}_j$ is a \(\mathcal P \mathcal M\) composed of two or more manifolds. To finalize the jet-level processing, all representations are mapped to the tangent space $\mathcal{T}_0 \mathcal{P}_j$ using the $log_0$ map and concatenated. Predictions are generated through a final Euclidean MLP followed by a \textsc{SoftMax} activation function.

\section{Product Manifold Representations for Jet Classification in JetClass}
\label{sec:perf_compare}
To compare the performance of variations in possible particle-level and jet-level representations, we train \(\mathcal P \mathcal M\)-MLP (\Cref{sec:PMNN}) and $\mathcal P \mathcal M$-Transformer (\Cref{sec:PMTrans}) models under several choices of \(\mathcal P \mathcal M\) representations for jet classification. 

\paragraph{Dataset}All experiments in this section utilize the open source \textsc{JetClass}\footnote{The dataset can be found here: \url{https://zenodo.org/records/6619768}} dataset \cite{qu_2022_particle,JetClass}.
The \textsc{JetClass} dataset consists of ten classes of simulated jets, represented as point clouds.
For the \(\mathcal P \mathcal M\)-MLP model we use the 20 highest $p_T$ particles in each jet and for the \(\mathcal P \mathcal M\)-Transformer for the 128 highest $p_T$ particles in each jet. 
Particle features consist of kinematic information (\(\Delta \eta\), \(\Delta \phi \), \(\log p_T\), \(\log E\), \(\log \frac{p_T}{p_T(\text{jet})}\), \(\log \frac{E}{E_\text{jet}}\), \( \Delta R \) ), particle identification, and trajectory displacement ($\tanh d_0, \tanh d_z$). Further details are listed in Table 2 of \cite{qu_2022_particle}.

\paragraph{\(\mathcal P \mathcal M\)-MLP Training Parameters}\(\mathcal P \mathcal M\)-MLP models in this section are trained with a batch size of 1024 for 20k batches, totaling over 5M samples. Training utilizes with a cosine annealing learning rate scheduler with an initial learning rate of \(10^{-3}\) which remains constant for 30\% of the iterations and then decays exponentially to 1\% of initial learning rate. We repeat all trainings over five initializations, presenting results and uncertainties for the highest accuracy models. We use the Riemannian Adam \cite{geoopt2020kochurov} with weight decay of 0, $\beta_1 = 0.9$, and $\beta_2 = 0.999$. \(\mathcal P \mathcal M\)-MLP models are all trained on a single A100 GPU. 

\paragraph{$\mathcal P \mathcal M$-Transformer Training Parameters}
$\mathcal P \mathcal M$-Transformer models in this section utilize 4 $\mathcal P \mathcal M$ Particle Attention blocks with 4 attention heads each. 
Models are trained with a batch size of 256 for 200k batches, totaling over 50M samples. When using particle-level representations of size 128D or larger, we train for 300k batches for convergence. Training utilizes with a cosine annealing learning rate scheduler with an initial learning rate of \(5\cdot 10^{-4}\) which remains constant for 30\% of the iterations and then decays exponentially to 1\% of initial learning rate. We repeat all trainings over five initializations, presenting results and uncertainties for the highest accuracy models. We use the Riemannian Adam with weight decay of 0, $\beta_1 = 0.9$, and $\beta_2 = 0.999$ \cite{geoopt2020kochurov}. \(\mathcal P \mathcal M\)-Transformer models are all trained in parallel on two or four A100 GPUs depending on model size.

\begin{figure}[h]
    \centering
    % First row with two images
    \begin{subfigure}[b]{0.49\linewidth}
       \includegraphics[width=\linewidth]{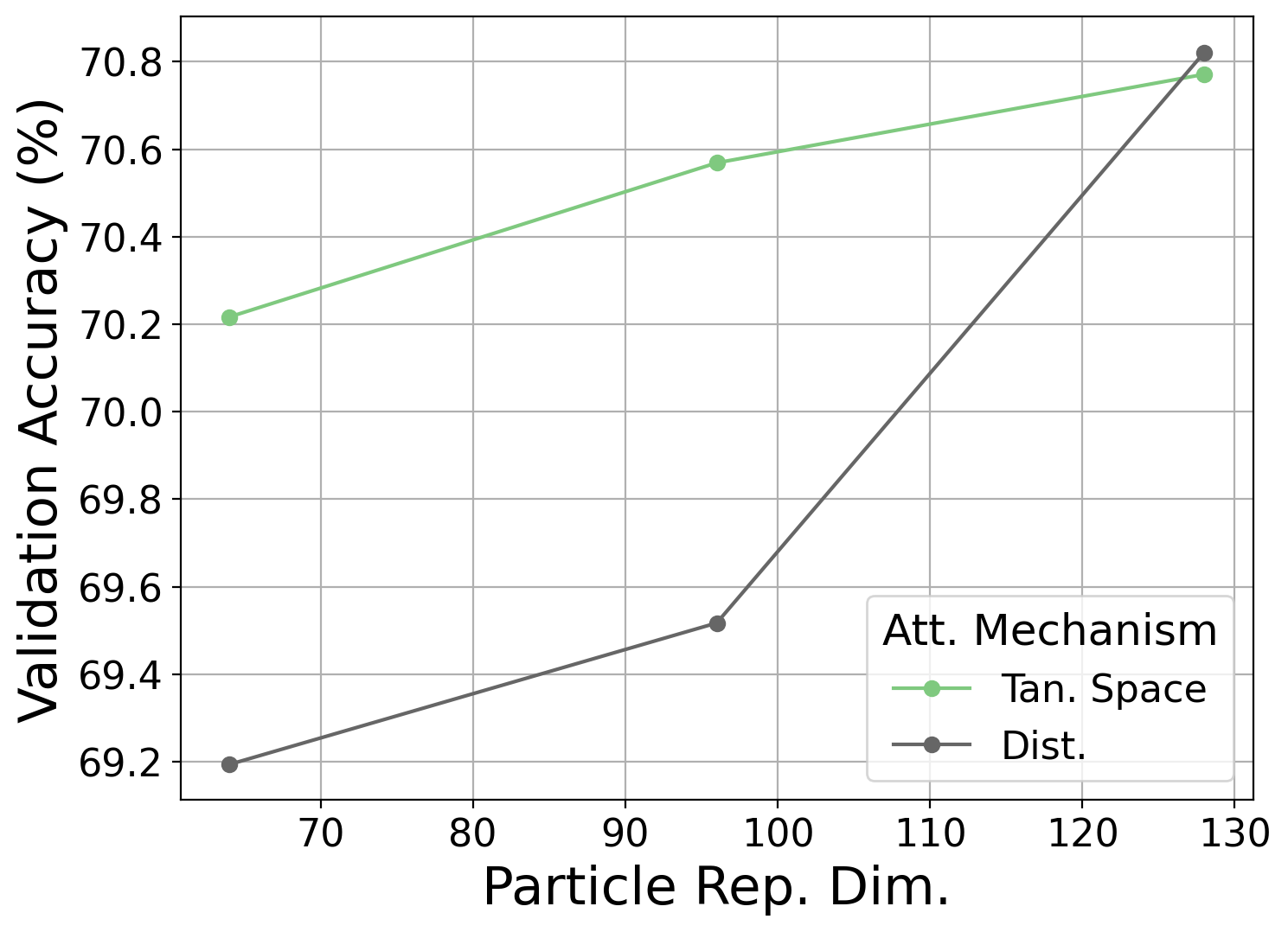}
        % Optional: overall caption for the figure
        \caption{Attention mechanism comparison on JetClass dataset.}
        \label{fig:att_mechanism_comparison}
    \end{subfigure}
    % \newline
    \hfill % adds horizontal space between the figures
    \begin{subfigure}[b]{0.49\linewidth}
        \includegraphics[width=\linewidth]{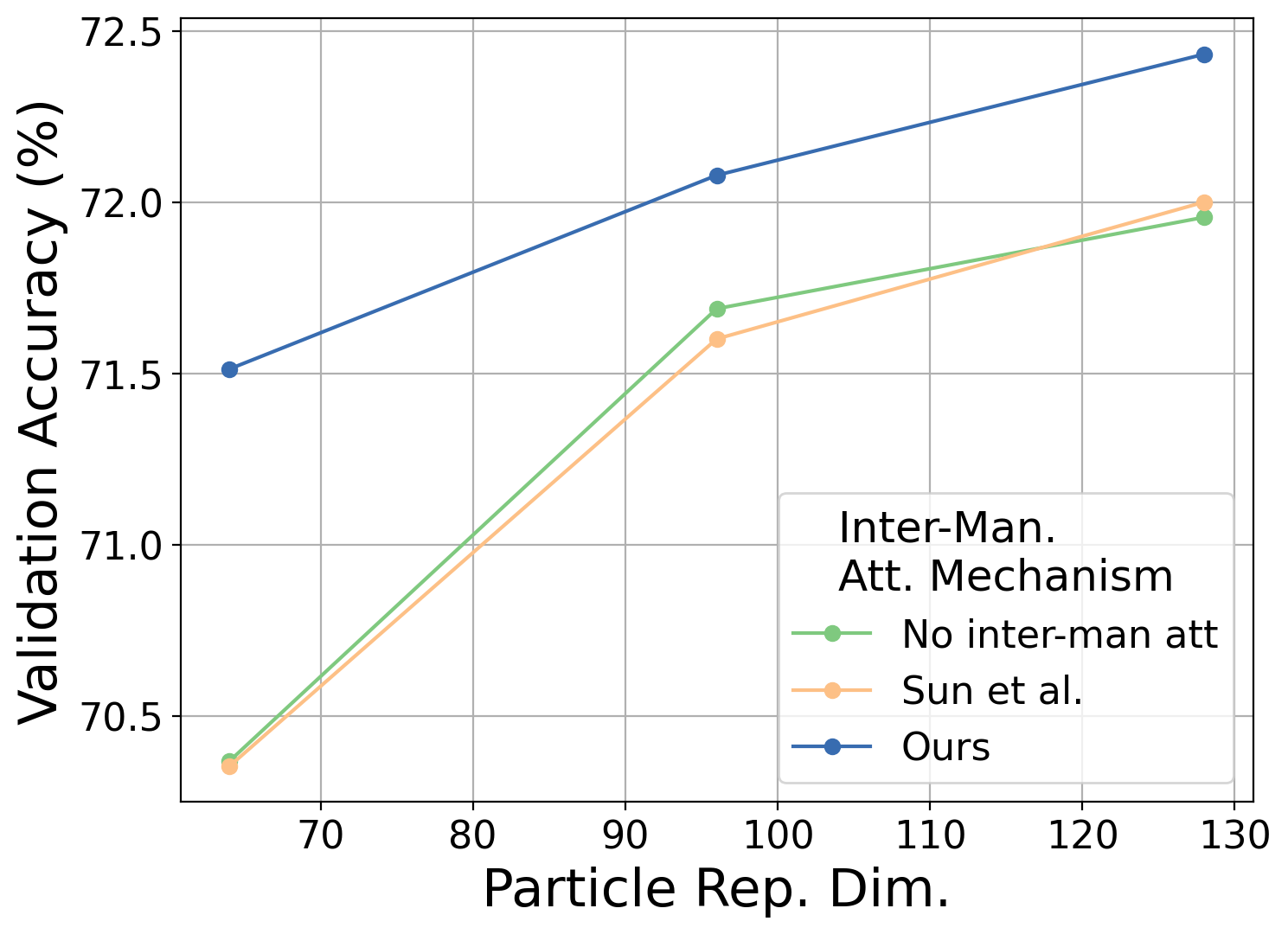}
        
        % Optional: overall caption for the figure
        \caption{Inter-manifold attention mechanism comparison on JetClass dataset.}
        \label{fig:inter_man_comparison}
    \end{subfigure}
    \caption{Performance comparison for architecture parameters in the \(\mathcal P \mathcal M\)-Transformer model}
    \label{fig:attention}
\end{figure}

\paragraph{Impact of Model Architecture on $\mathcal P \mathcal M$-Transformer Performance} 
As stated in \Cref{sec:manifold_ml}, for attention mechanisms and inter-manifold attention we propose alternative approaches from that of the literature. To highlight the impact of our proposed methods, we perform comparison studies on the \textsc{JetClass} multi-class  classification tasks. For all studies shown in \Cref{fig:attention}, we use the $\mathcal P \mathcal M$-Trans($\mathcal P_p =\mathbb R^{32} \times \mathbb{H}^{32}$,$\mathcal P_j = \mathbb R^{16}$) with training parameters described above. 

We compare distance-based attention \cite{aglarglehre_2018_hyperbolic} vs attention through inner products in the tangent space in \Cref{fig:att_mechanism_comparison}, finding gains through tangent space attention. For inter-manifold attention, we compare three methods in \Cref{fig:inter_man_comparison}: no inter-manifold attention, Sun et al.'s \cite{sun_2022_a} original method, and our proposed method. We find our novel method of inter-manifold attention to perform the best in this comparison. 
To better understand the impact of model features on performance, we perform an ablation study shown in \Cref{tab:accuracy_table}. The base model only contains \textsc{ReLU} activations in Euclidean spaces. We find slight performance gains through introducing \textsc{ReLU} activations in non-Euclidean spaces; aligning with our understanding that non-Euclidean spaces are innately non-linear and do not necessitate the injection of non-linearity. Adding \textsc{LayerNorm} operations in the tangent space and inter-manifold attention between representations both yield relatively large performance increases.

\begin{table}[htbp]
\centering
\begin{tabular}{lc}
\hline
\textbf{Model/Feature} & \textbf{Accuracy} \\
\toprule
Base Model: $\mathcal P \mathcal M$-Trans($\mathcal P_p =\mathbb R^{32} \times \mathbb{H}^{32}$,$\mathcal P_j = \mathbb R^{16}$) & 68.70 \\

+ \textsc{ReLU} activation in Non-Euclidean MLPs & 68.83 $\uparrow$ 0.13 \\ 

+ Tangent space \textsc{LayerNorm} & 69.64 $\uparrow$ 0.81\\

+ inter-manifold attention & 70.67 $\uparrow$ 1.03 \\
\bottomrule

\end{tabular}
\caption{Ablation study for $\mathcal P \mathcal M$-Transformer model architecture.}
\label{tab:accuracy_table}
\end{table}
\vspace{2em}
\subsection{Impact of Product Manifold Representations at the Particle-Level}
\begin{wrapfigure}{r}{0.65\textwidth}
\vspace{-0.5cm}
\includegraphics[width=\linewidth]{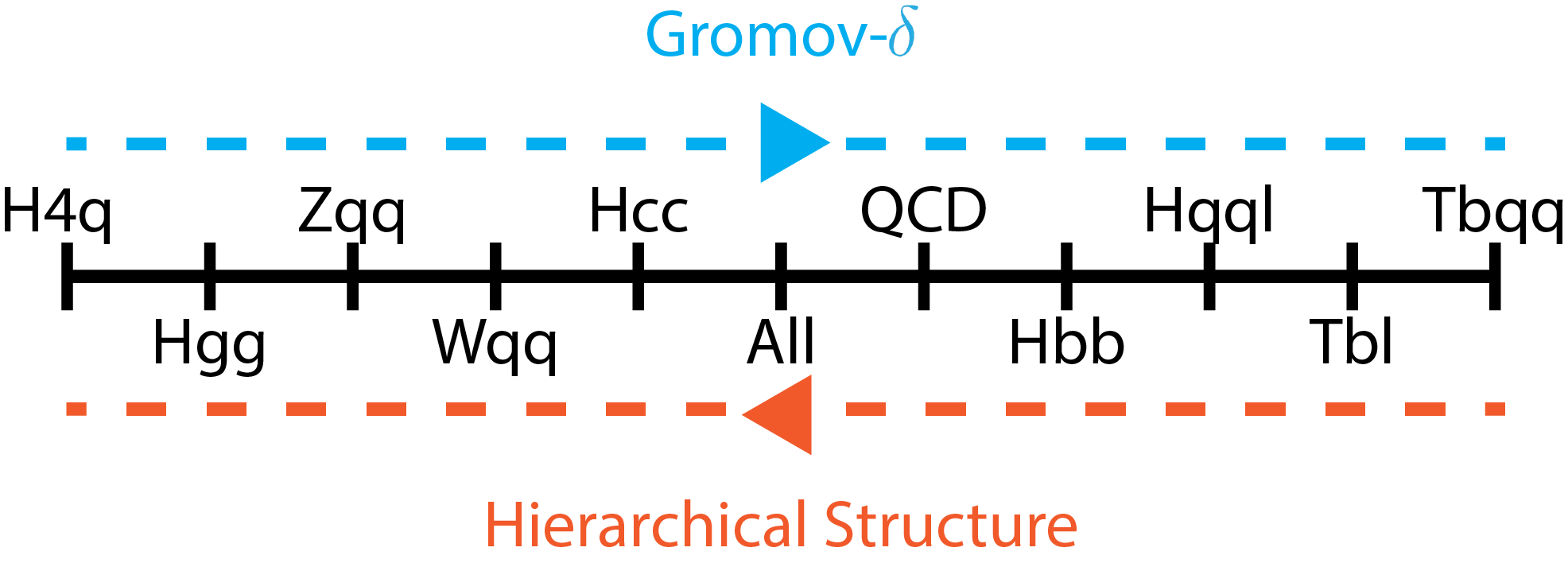}
    \caption{Graphic illustrating the order of Gromov-$\delta$ hyperbolicity estimate for all classes in JetClass. `All' represents the hyperbolicity computed over all classes combined. We find H4q has the most hierarchical structure and Tbqq has the least hierarchical structure.}
    \label{fig:gromov_d}
    \vspace{-3em}
\end{wrapfigure}
\label{sec:part_lvl}
As described in \Cref{sec:intro}, due to showering and hadronization in jet formation, we expect jet constituents to have a hierarchical relationship. 
To further understand the hierarchical structure of the ten jet classes, we estimate the relative Gromov-\(\delta\) hyperbolicity (see \Cref{sec:gromov}) for each class by averaging the hyperbolicity estimates over random samples of jets within each class.
The relative ordering\footnote{While the Gromov-$\delta$ hyperbolicity is useful as it can be estimated numerically, we are not aware of a well defined method to calculate a curvature with certainty. As such, we solely provide the relative ordering of hyperbolicity.} is shown in \Cref{fig:gromov_d}. 

\subsubsection{Product Manifold Neural Network}
\label{sec:part_PM-NN}
\begin{figure}[h!]
    \centering
    % First row with accuracy plots
    \begin{subfigure}[b]{0.49\linewidth}
        \centering
        \includegraphics[width=\linewidth]{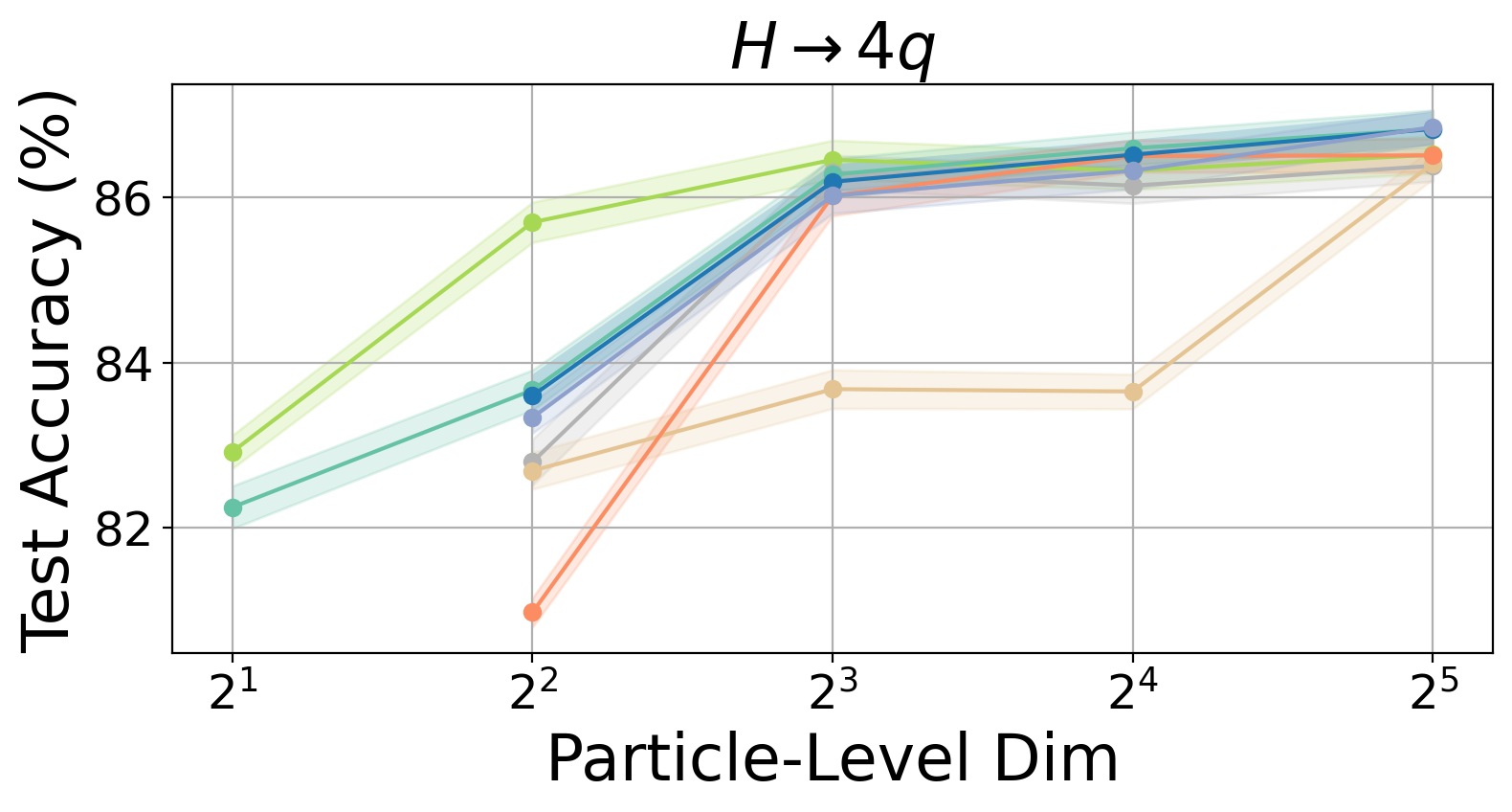}
        \caption{Accuracy for $H \to 4q$}
        \label{fig:h4q_acc}
    \end{subfigure} 
    \hfill
    \begin{subfigure}[b]{0.49\linewidth}
        \centering
        \includegraphics[width=\linewidth]{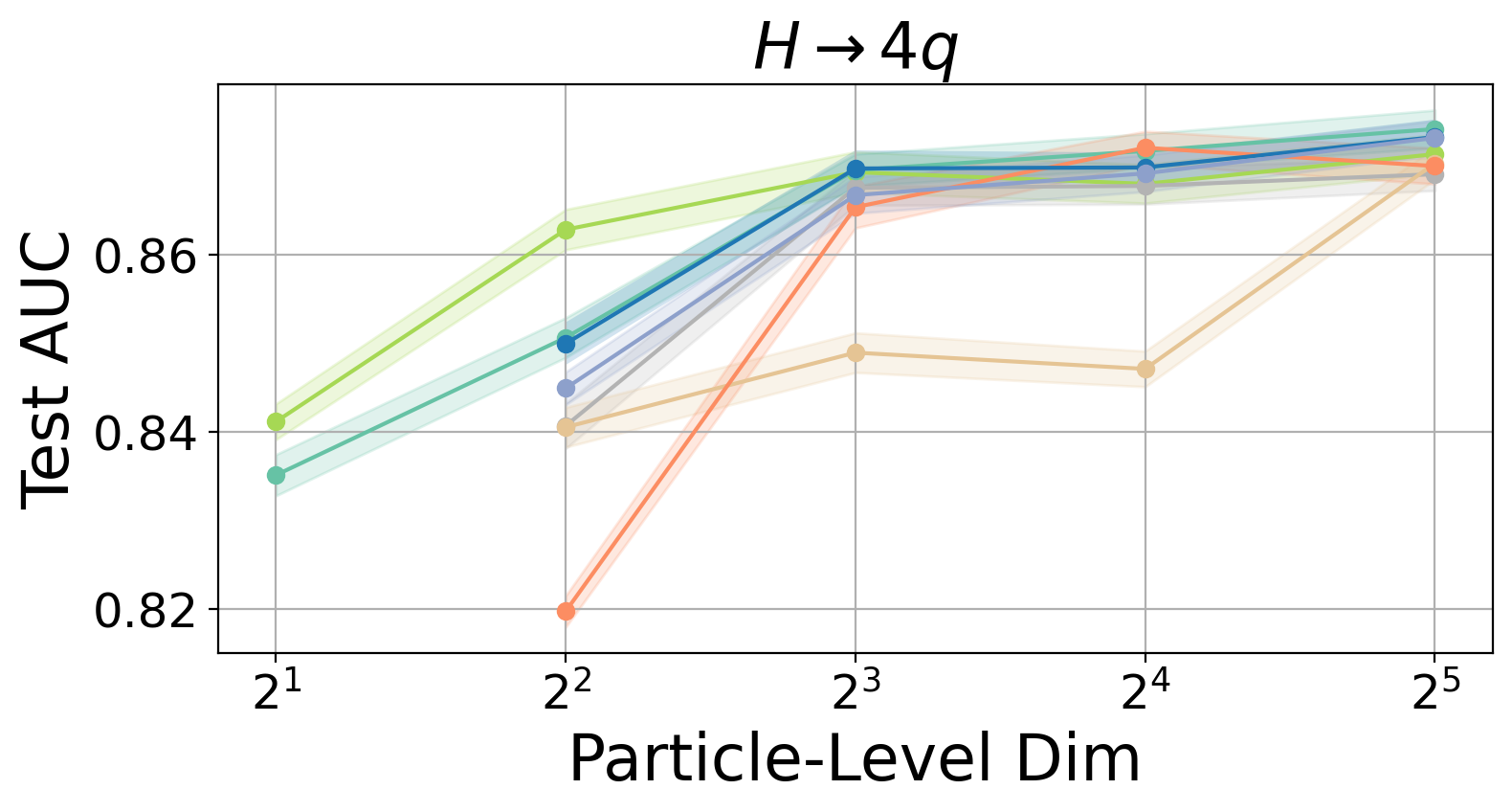}
        \caption{AUC for $H \to 4q$}
        \label{fig:h4q_auc}
    \end{subfigure} 
    % First row with accuracy plots
    \begin{subfigure}[b]{0.49\linewidth}
        \centering
        \includegraphics[width=\linewidth]{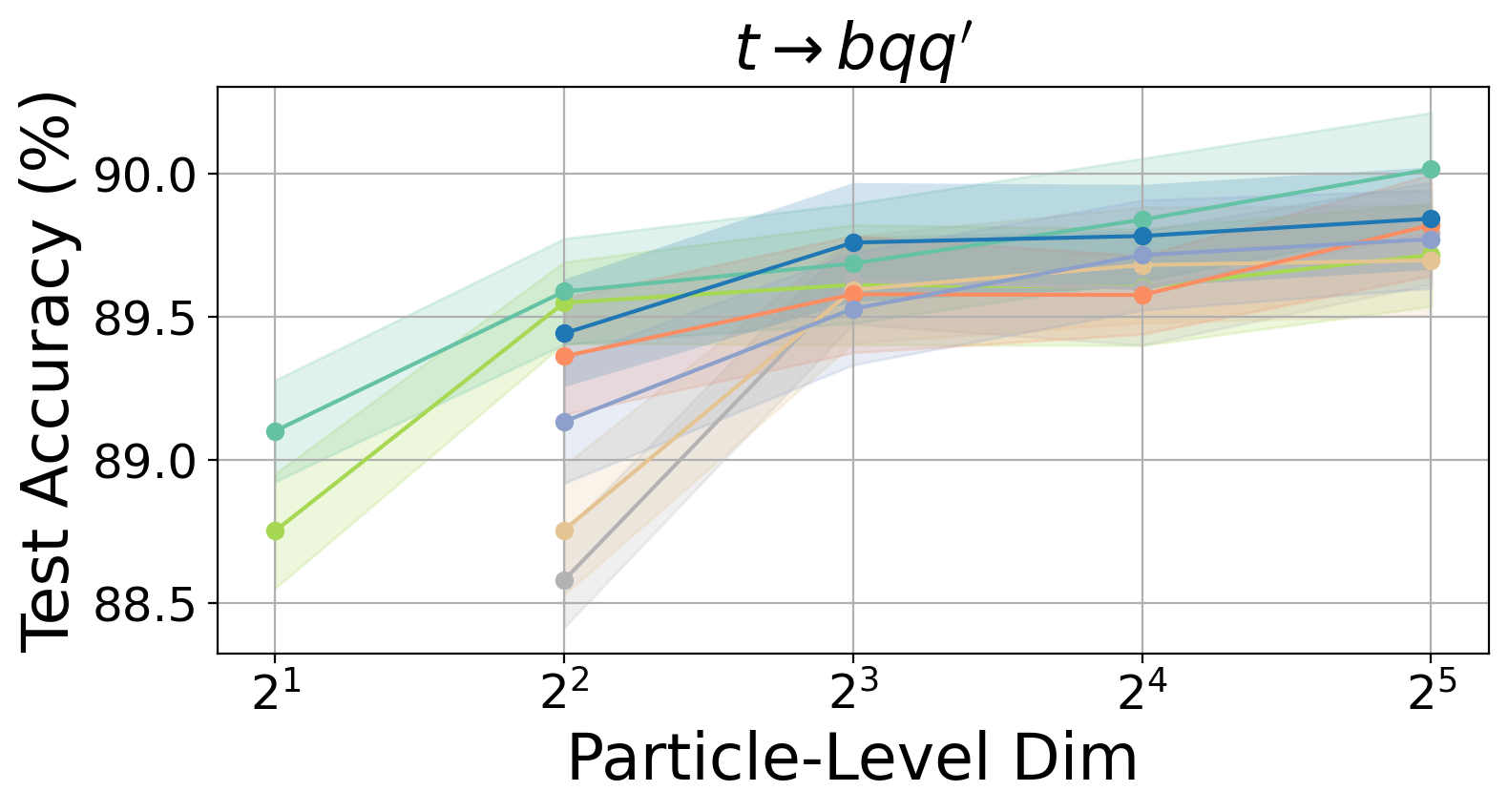}
        \caption{Accuracy $t \to bqq'$}
        \label{fig:tbqq_acc}
    \end{subfigure}
    \hfill
    \begin{subfigure}[b]{0.49\linewidth}
        \centering
        \includegraphics[width=\linewidth]{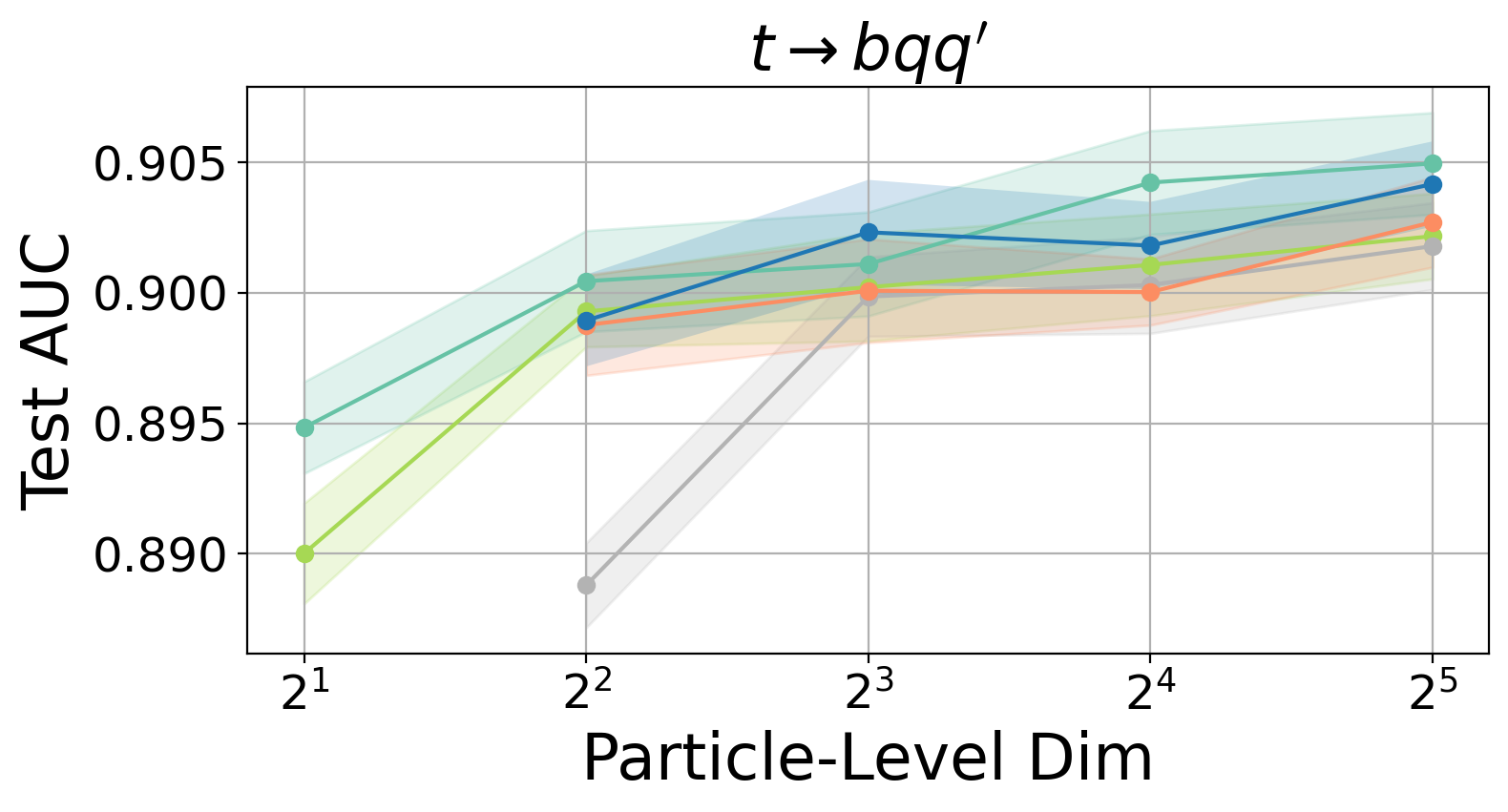}
        \caption{AUC $t \to bqq'$}
        \label{fig:tbqq_auc}
    \end{subfigure} 
    
    \begin{subfigure}[b]{\linewidth}
        \centering
    \includegraphics[width=0.9\linewidth]{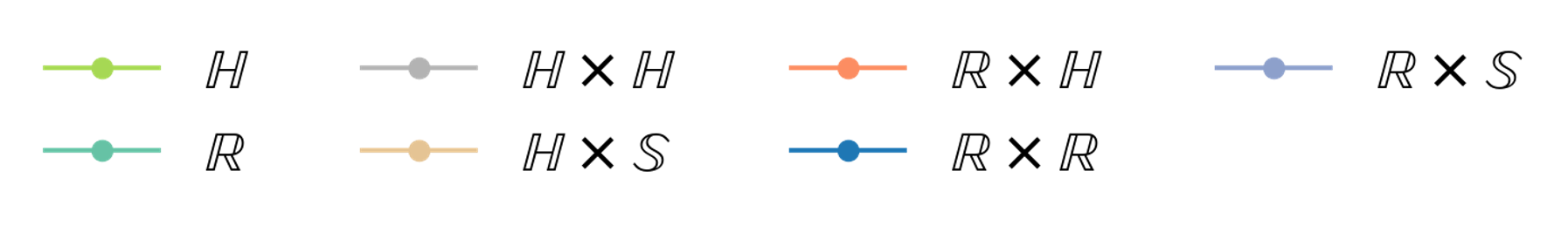}
    \end{subfigure} 
    \caption{\(\mathcal{P}\mathcal{M}\)-MLP performance on $H \to 4q$ vs QCD and $t \to bqq'$ vs QCD.}
    \label{fig:pm_MLP_results}
\end{figure}

We evaluate the performance of \(\mathcal P \mathcal M\)-MLP models in binary classification of $H \to 4q$ vs QCD and $t \to bqq'$ vs QCD using several particle-level geometries of total dimension 4D to 32D. We choose these signals based on the relative values of Gromov-hyperbolicity shown in \Cref{fig:gromov_d}, where $H \to 4q$ is the most hierarchical process and $t \to bqq'$ is the least hierarchical process. Results for both sets of experiments are combined and presented in \Cref{fig:pm_MLP_results}.

From \Cref{fig:pm_MLP_results}, for highly hierarchical classes (\(H \rightarrow 4q\), \Cref{fig:h4q_acc,fig:h4q_auc}), hyperbolic representations outperforms Euclidean representations at dimensions 2 and 4, while achieving comparable performance at higher dimensions. For weakly hierarchical classes (\(t \rightarrow bqq'\), \Cref{fig:tbqq_acc,fig:tbqq_auc}), the performance of \(\mathcal{P}\mathcal{M}\) representations is generally within the uncertainty range of fully Euclidean representations.
To analyze the structure of these embeddings, we present the tangent space corner plots of particle embeddings for \(\mathcal{P}\mathcal{M}\)-MLP models with \(\mathbb{R}^4\) and \(\mathbb{H}^4\) geometries in Appendix \ref{sec:plvl_embed}.

This performance aligns with expectations based on the hierarchical structure inherent in the dataset, illustrating that the Gromov-\(\delta\) framework effectively underscores the advantages of hyperbolic representations in enhancing model performance. In both processes, combining Euclidean and non-Euclidean representations does not yield performance gains for \(\mathcal P \mathcal M\)-MLP models. This suggests that low-parameter model and simple  architectures may not be able to effectively leverage the additional perspectives through parallel representations.

To further explore the use of \(\mathcal P \mathcal M\) representations at the particle-level and realistically test their capabilities for deployment in jet analysis, we test the performance of the $\mathcal P \mathcal M$ transformer model (\Cref{sec:PMTrans}).

\subsubsection{Product Manifold Transformer Models}
\label{sec:part_PMtrans}
In this section, we consider particle-level representations of total dimension 48D to 240D (6D to 30D per attention head)  for \( \mathcal P \mathcal M\) representations  and 32D to 160D (8D to 40D per attention head) for fully Euclidean models. While these dimension ranges vary, the total parameters are similar for these models. The jet-level representation is fixed at \(\mathbb{R}^{16}\) to isolate the impact of particle-level representations on performance. We compare the performance in jet classification of all processes in the \texttt{JetClass} dataset across various of particle-level representations. We refer to models by their particle-level representation in all figures in this section.  

\begin{figure}[H]
    \centering
    % First row with accuracy plots
    \begin{subfigure}[b]{0.49\linewidth}
        \centering
        \includegraphics[width=\linewidth]{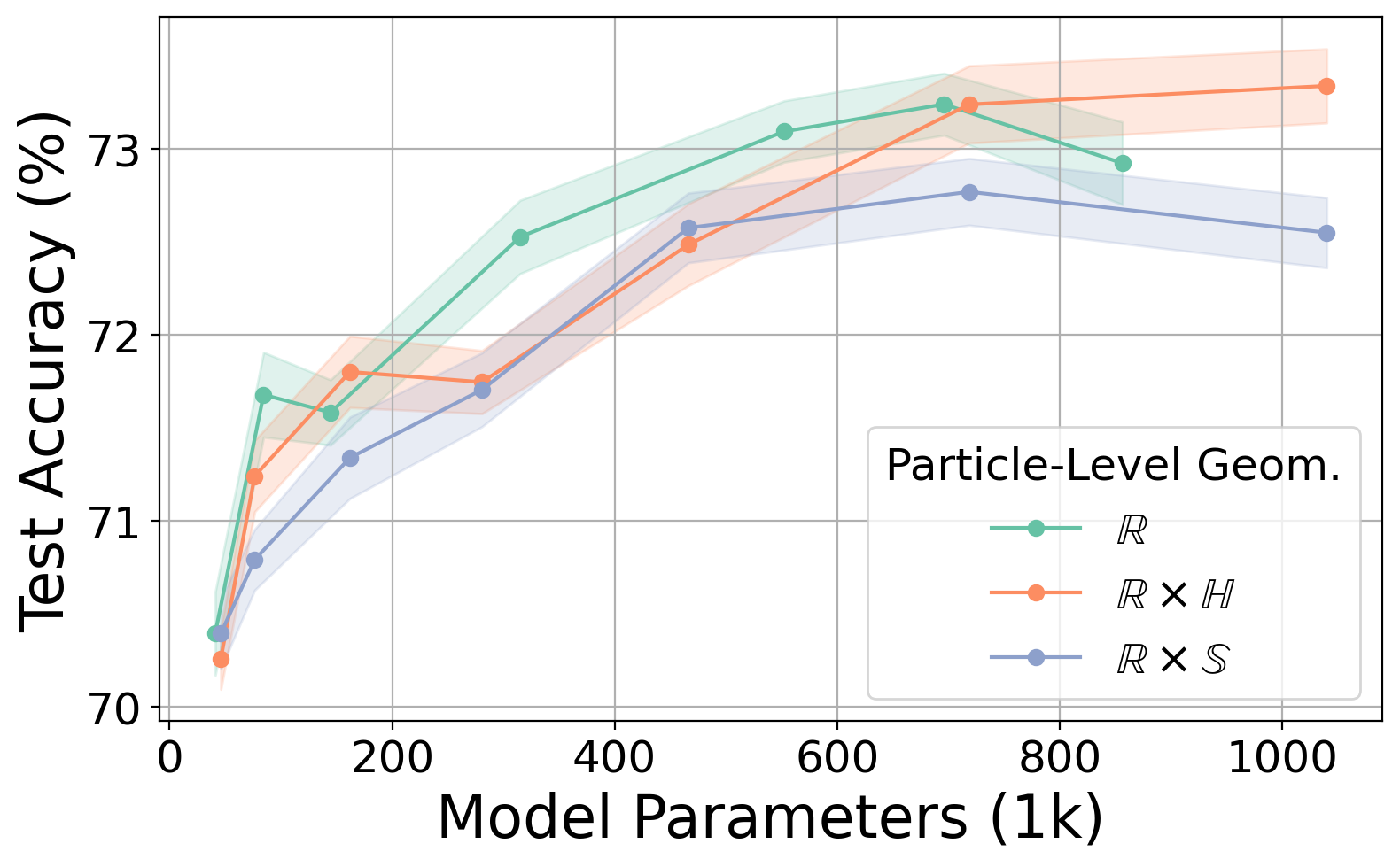}
        \caption{Accuracy for Particle-Level Representations}
        \label{fig:trans_part_lvl_acc}
        
    \end{subfigure} 
    \hfill
    \begin{subfigure}[b]{0.49\linewidth}
        \centering
        \includegraphics[width=\linewidth]{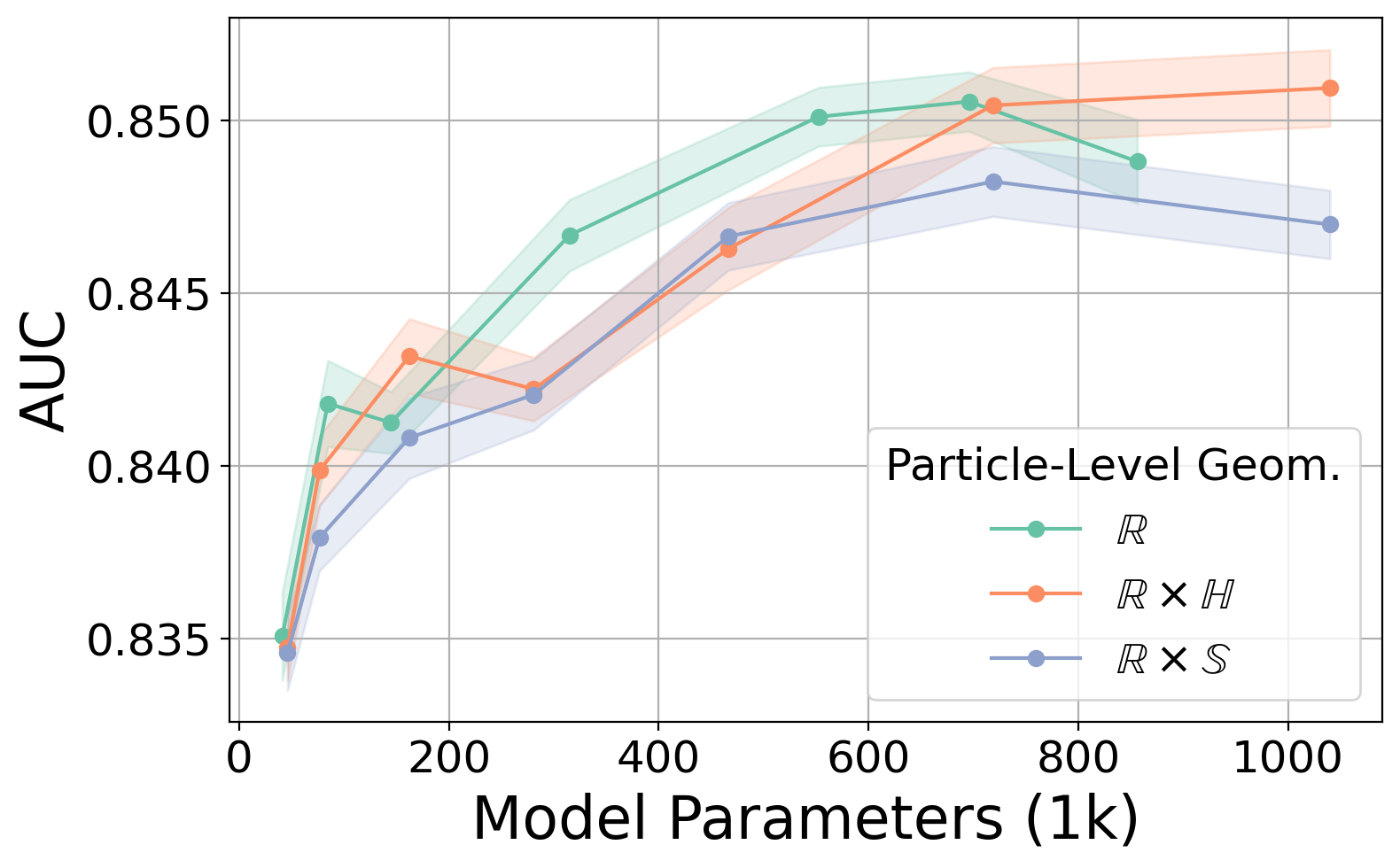}
        \caption{AUC for Particle-Level Representations}
        \label{fig:trans_part_lvl_auc}
    \end{subfigure} 
    
     \caption{Test Accuracy for several possible particle-level geometries in the $\mathcal P \mathcal M$-Transformer model. Models employing $\mathbb R \times \mathbb H$ performs equal or better than fully Euclidean counter parts for low-parameter and high-parameter models. For parameter ranges 300k to 500k, we find performance losses with \(\mathcal P \mathcal M\) representations.}
   
\end{figure}

\Cref{fig:trans_part_lvl_acc,fig:trans_part_lvl_auc} shows the test accuracies and AUC, respectively, for select particle-level geometries across models ranging from 70k to 1M parameters. Among the non-Euclidean approaches, \(\mathbb{R}\times\mathbb{H}\) representations perform the best. Compared to fully Euclidean models, \(\mathcal{P}\mathcal{M}\) representations exhibit similar accuracy at low parameter counts, a slight dip at intermediate parameter ranges, and potential improvements for larger models.
These results demonstrate that \(\mathcal{P}\mathcal{M}\) representations can be effectively integrated into models while maintaining, or even improving, performance in jet classification tasks. While the \texttt{JetClass} dataset is a common baseline, the ten classes with varying degrees of hierarchical structure make it a more challenging environment to analyze and interpret our model's performance. To address this, we also benchmark our models on binary classification tasks in \Cref{sec:bench}, where the controlled settings provide clearer insights into model behavior and effectiveness.

\subsection{Impact of Product Manifold Representations at the Jet-Level in the \(\mathcal P \mathcal M\)-Transformer}

\label{sec:jet_classification} 
In this section, we consider jet-level representations of total dimension 16D to 64D for all models. The particle-level representation is fixed at \(\mathbb{R}^{32}\) to isolate the impact of particle-level representations on performance. We compare the performance in jet classification of all processes in the JetClass dataset across various of jet-level representations. We refer to models by their jet-level representation in all figures in this section.  

\begin{figure}[H]
    \centering
    % First row with accuracy plots
    \begin{subfigure}[b]{0.49\linewidth}
        \centering
        \includegraphics[width=\linewidth]{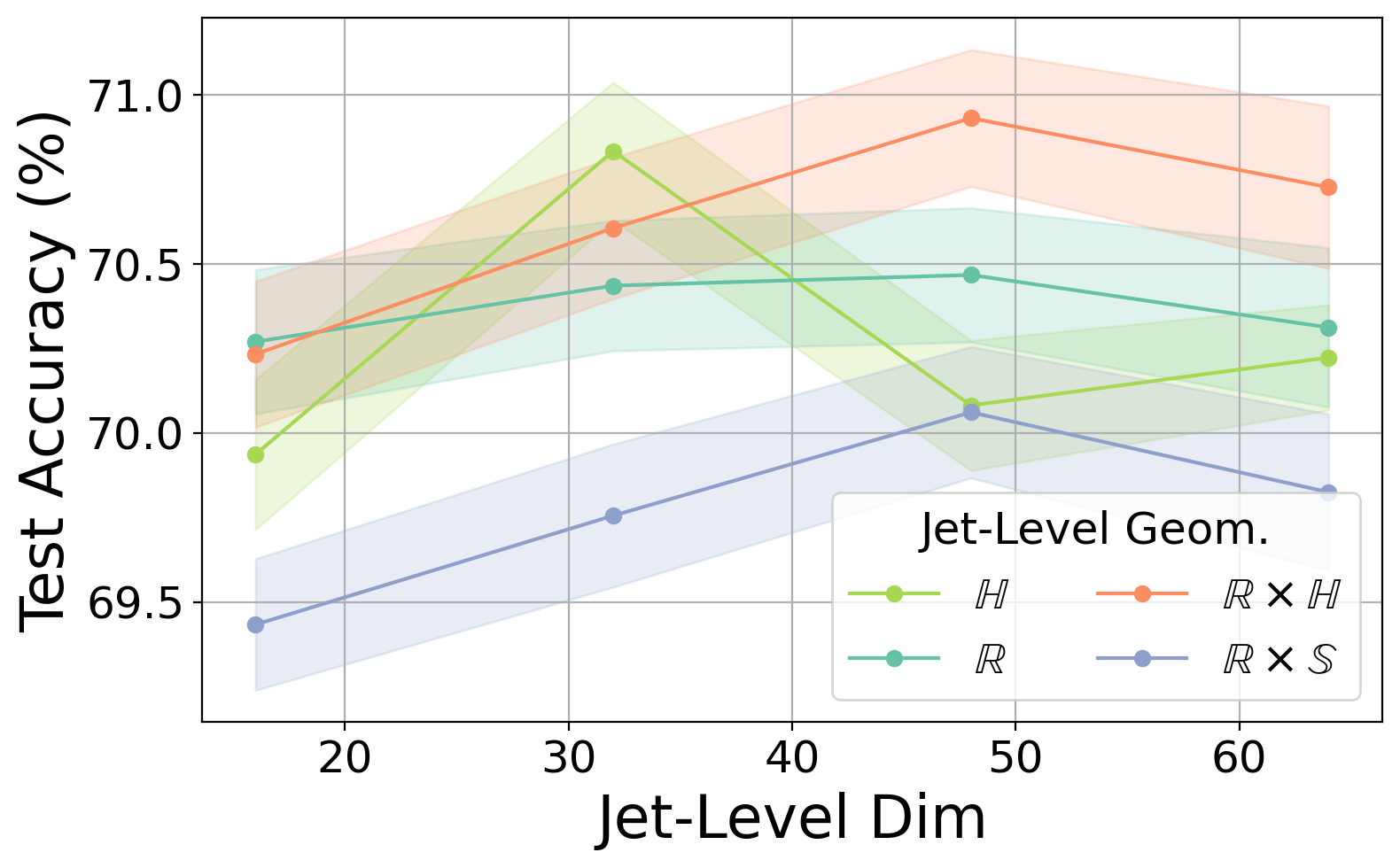}
        \caption{Accuracy for Jet-Level Representations}
    \label{fig:trans_jet_lvl_acc}
        
    \end{subfigure} 
    \hfill
    \begin{subfigure}[b]{0.49\linewidth}
        \centering
        \includegraphics[width=\linewidth]{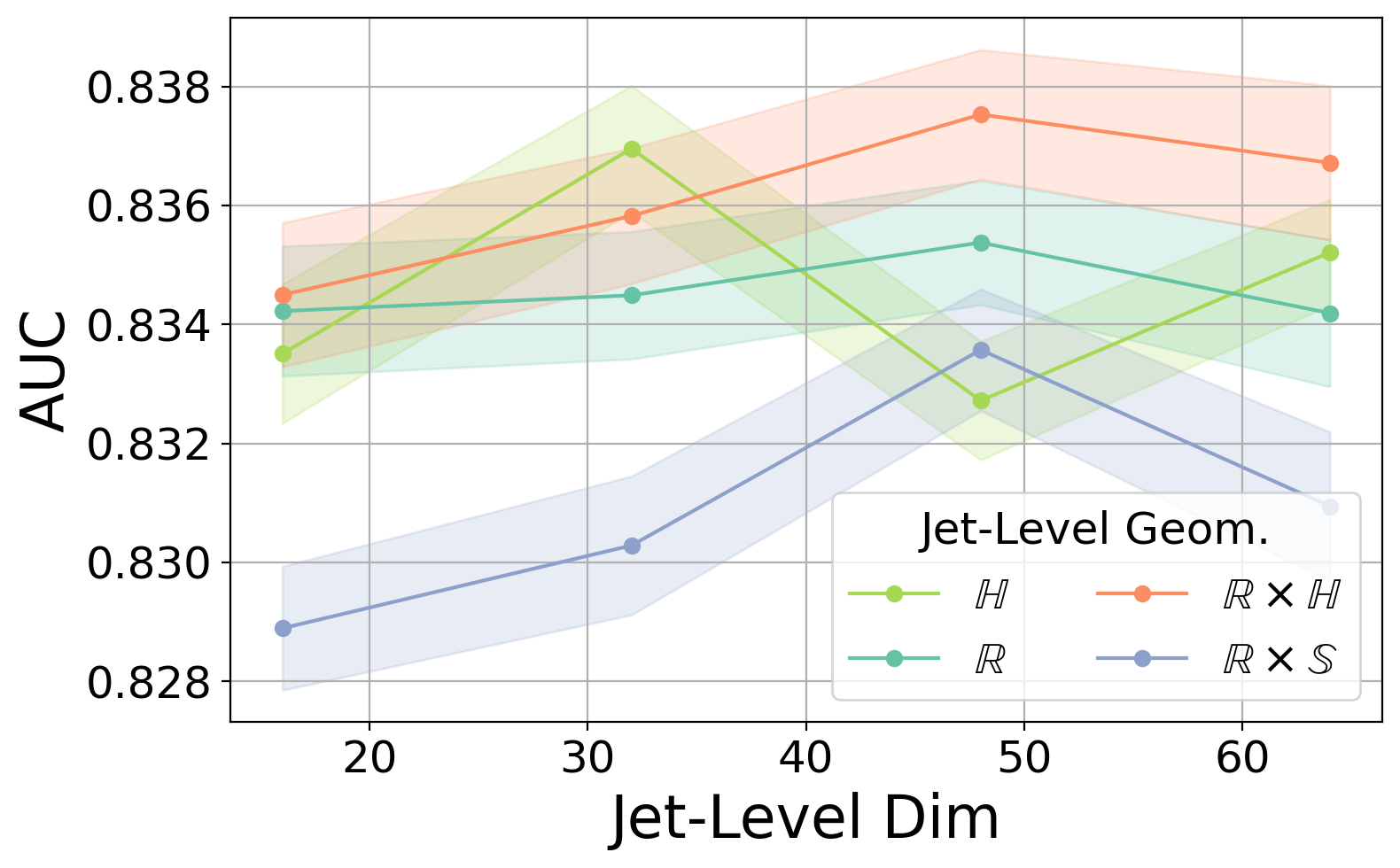}
        \caption{AUC for Jet-Level Representations}
        \label{fig:trans_jet_lvl_auc}
    \end{subfigure} 
    
     \caption{Test Accuracy for several jet-level geometries in the $\mathcal P \mathcal M$-Transformer model. All jet-level representation considered converge within error. When considering the best performing models from each category, \(\mathbb R \times \mathbb H\) achieves the highest performance considering. This suggests that \(\mathbb R \times \mathbb H\) for jet-level representations can bring equal or greater performance compared to fully Euclidean representations.}
   
\end{figure}

We find that \(\mathcal P \mathcal M\) representations at the jet-level (\Cref{fig:trans_jet_lvl_acc,fig:trans_jet_lvl_auc} are largely inconsequential in classification tasks over the range of jet-level representation dimension considered. While all models converge largely within error, we can see the best performing models at each dimension are predominately non-Euclidean. 
Specifically, \(\mathbb R \times \mathbb H\) consistently achieves equal or better performance compared to fully Euclidean representations when considering the best models.   
This indicates potential advantages in utilizing \(\mathcal{P} \mathcal{M}\) representations at the jet level with hyperbolic embeddings for classification tasks, though definitive conclusions cannot yet be drawn. 
However, we emphasize that jet-level representations are unlikely to remain insignificant for other tasks, such as embedding learning, as suggested by \cite{sangeonpark_2023_neural}.

\section{Performance Benchmarks}
\label{sec:bench}
Thus far we have compared the performance effects on JetClass classification of possible particle-level and jet-level representations within the \(\mathcal{P} \mathcal{M}\)-MLP (\Cref{sec:part_PM-NN}) and \(\mathcal{P} \mathcal{M}\) transformer (\Cref{sec:part_PMtrans,sec:jet_classification}) and found that \(\mathcal{P} \mathcal{M}\) representations can bring performance gains. 
To further explore the performance of \(\mathcal{P} \mathcal{M}\) models, we benchmark the \(\mathcal{P} \mathcal{M}\)-Transformer model on top tagging. 
The classification of top jets versus QCD background is a widely explored benchmark for ML-based jet taggers in particle physics. As such, we select top tagging as a first comparison to existing literature \cite{tagger_bogatskiy2020lorentzgroupequivariantneural,tagger_das2022featureselectiondistancecorrelation,tagger_Gong_2022,tagger_Kasieczka_2019,tagger_Komiske_2019,tagger_Munoz_2022,tagger_pearkes2017jetconstituentsdeepneural} for the \(\mathcal P \mathcal M\)-Transformer model. 
For all models, we use the \(\mathbb R \times \mathbb H\) representation at the particle-level and jet-level. 

\subsection{Top Tagging}
The top tagging benchmark dataset \cite{kasieczka_gregor_2019_2603256,Kasieczka:2019dbj} provides 1.2M training, 400k validation and 400k test data samples for top ($t \rightarrow bqq'$) vs QCD ($g/q$) classification. Particle features are restricted to kinematic information  ($\Delta \eta, \Delta \phi, \log p_T, \log E, \log \frac{p_T}{p_T(\text{jet})}, \log \frac{E}{E_\text{jet}}, \Delta R$). 
We consider $\mathcal P \mathcal M$-Transformer models with total parameters ranging from 664 to 858k. 
All models are trained for 30 epochs with a batch size of 128 over the entire training dataset. Training utilizes with a cosine annealing learning rate scheduler with an initial learning rate of \(5 \cdot 10^{-4}\) which remains constant for 50\% of the iterations and then decays exponentially to 1\% of initial learning rate. We repeat all trainings over three initializations, presenting results and uncertainties for the best performing models.

\begin{figure}[h]
    \centering
    \includegraphics[width=\linewidth]{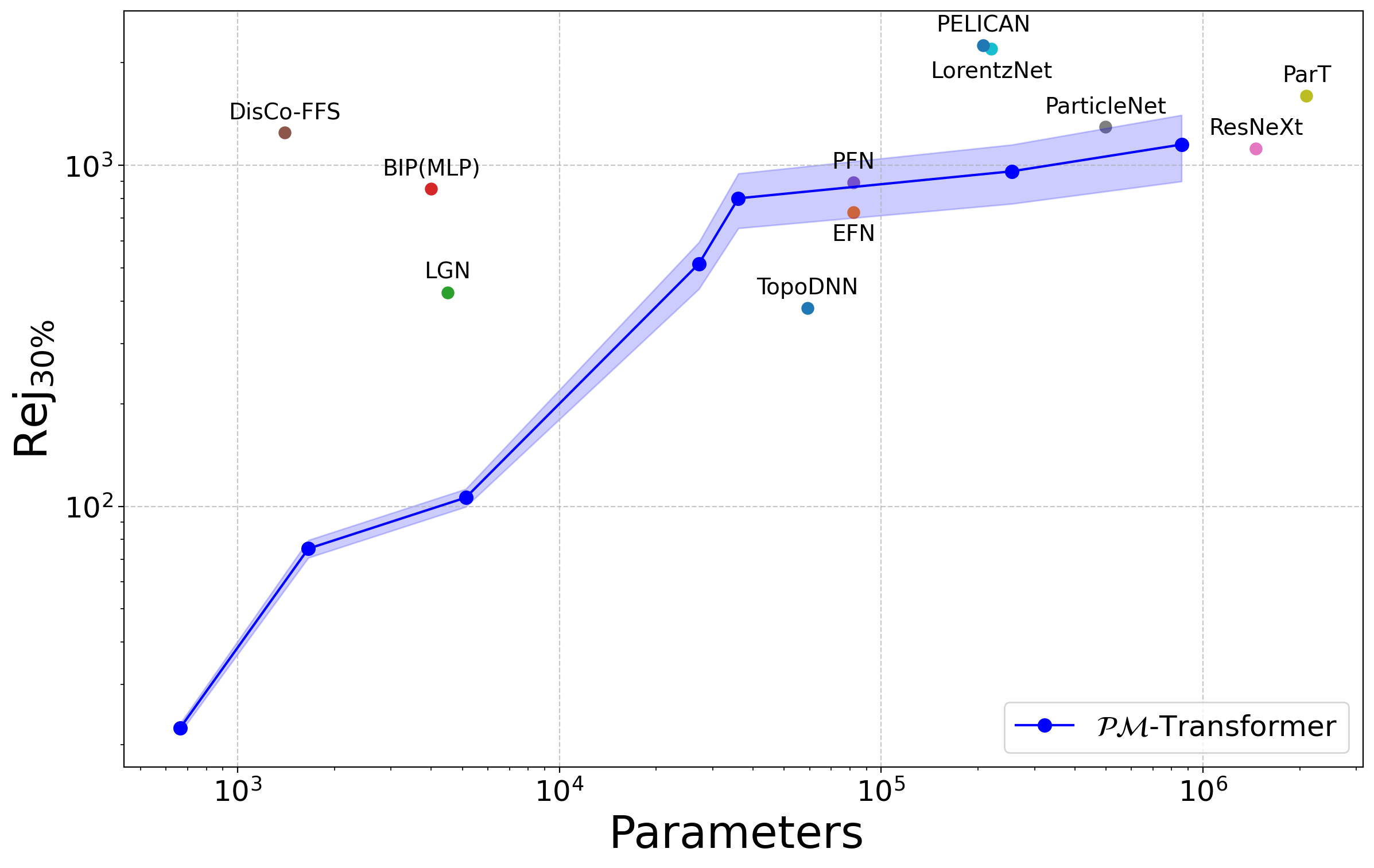}
    \caption{Performance of $\mathcal P \mathcal M$-Transformer in parameter space relative to other top taggers \cite{tagger_bogatskiy2020lorentzgroupequivariantneural,tagger_das2022featureselectiondistancecorrelation,tagger_Gong_2022,tagger_Kasieczka_2019,tagger_Komiske_2019,tagger_Munoz_2022,tagger_pearkes2017jetconstituentsdeepneural}. We only compare to models trained exclusively on the training dataset provided in the top tagging benchmark dataset \cite{kasieczka_gregor_2019_2603256,Kasieczka:2019dbj}.}
    \label{fig:top_rej}
\end{figure}

\begin{table}[h]
    \centering
    \begin{tabular}{lccccc}
        \toprule
        & Parameters & Accuracy & AUC & Rej$_{50\%}$ & Rej$_{30\%}$ \\
        \midrule
        $\mathcal P \mathcal M$-Transformer & 858k & 93.8 & 0.9842 & $295 \pm 33$ & $1151 \pm 253$ \\
        ParT & 2.1M & 94.0 & 0.9858 & $413 \pm 16$ & $1602 \pm 81$ \\
        \bottomrule
    \end{tabular}
    \caption{Comparison of the best performing $\mathcal P \mathcal M$-Transformer with ParT \cite{qu_2022_particle}, another transformer-based model.}
    \label{tab:top_compare}
\end{table}
\vspace{-1em}
The best performing $\mathcal P \mathcal M$-Transformer achieves an overall accuracy of 93.8\% on the test dataset with AUC and rejection at 30\% and 50\% listed in \Cref{tab:top_compare}. Compared to the ParT model \cite{qu_2022_particle}, another transformer-based particle tagger, we achieve performance within 0.2\% while using 60\% less parameters. To further understand the performance of the $\mathcal P \mathcal M$-Transformer over the full parameter space, we train several lower parameter configurations. We plot model performance relative to the number of trainable parameters in \Cref{fig:top_rej} for models with 664 to 858k parameters. Further information on these models is provided in \Cref{tab:pm_top_scaling}.

\begin{table}[ht]
\centering
\scalebox{0.95}{
\begin{tabular}{@{}cccccccc@{}}
\toprule
\textbf{Params} & \textbf{Accuracy (\%)} & \textbf{AUC} & $\textbf{Rej}_{30\%}$ & \textbf{Particle Geom} & \textbf{Jet Geom} & \textbf{Layers} & \textbf{Heads} \\ \midrule
664    & $86.2 \pm 0.2$ & 0.909 & $22 \pm 1$    & $\mathbb{R}^2 \times \mathbb{H}^2$   & $\mathbb{R}^2 \times \mathbb{H}^2$   & 1  & 1  \\
1.66k   & $89.9 \pm 0.1$ & 0.952 & $75 \pm 4$    & $\mathbb{R}^4 \times \mathbb{H}^4$   & $\mathbb{R}^4 \times \mathbb{H}^4$   & 1  & 1  \\
5.12k   & $90.3 \pm 0.2$ & 0.958 & $106 \pm 7$   & $\mathbb{R}^5 \times \mathbb{H}^{5}$ & $\mathbb{R}^{12} \times \mathbb{H}^{12}$ & 2  & 1  \\
27.16k  & $92.7 \pm 0.1$ & 0.978 & $515 \pm 81$  & $\mathbb{R}^{10} \times \mathbb{H}^{10}$ & $\mathbb{R}^{24} \times \mathbb{H}^{24}$ & 2  & 1  \\
36.01k  & $93.1 \pm 0.1$ & 0.981 & $801 \pm 146$ & $\mathbb{R}^{10} \times \mathbb{H}^{10}$ & $\mathbb{R}^{24} \times \mathbb{H}^{24}$ & 8  & 1  \\
254.55k & $93.4 \pm 0.1$ & 0.983 & $961 \pm 189$ & $\mathbb{R}^{40} \times \mathbb{H}^{40}$ & $\mathbb{R}^{48} \times \mathbb{H}^{48}$ & 8  & 4  \\
858.19k & $93.8 \pm 0.1$ & 0.984 & $1152 \pm 254$ & $\mathbb{R}^{80} \times \mathbb{H}^{80}$ & $\mathbb{R}^{64} \times \mathbb{H}^{64}$ & 8  & 8  \\
 \bottomrule
\end{tabular}
}
\caption{Performance of $\mathcal P \mathcal M$-Transformer Models at Varying Parameters. Layers refers to the number of $\mathcal P \mathcal M$ particle attention block. All models have two $\mathcal P \mathcal M$ class attention blocks}
\label{tab:pm_top_scaling}
\end{table}

\subsection{Performance Correlation with Gromov-$\delta$ Hyperbolicity}
\label{sec:gromov_binning}

For the \(\mathcal P \mathcal M\)-MLP model (\Cref{sec:PMNN}) we found that hyperbolic representations can bring performance gains in particle jet processes with a high degree of hierarchical structure (\Cref{fig:pm_MLP_results}). To further understand the performance of the \(\mathcal P \mathcal M\)-Transformer model, we determine model performance on the top tagging dataset relative to the calculated Gromov-\(\delta\) hyperbolicity per jet. 

\begin{wrapfigure}{r}{0.60\textwidth}
    \centering
    \vspace{-0.5cm}
    \includegraphics[width=\linewidth]{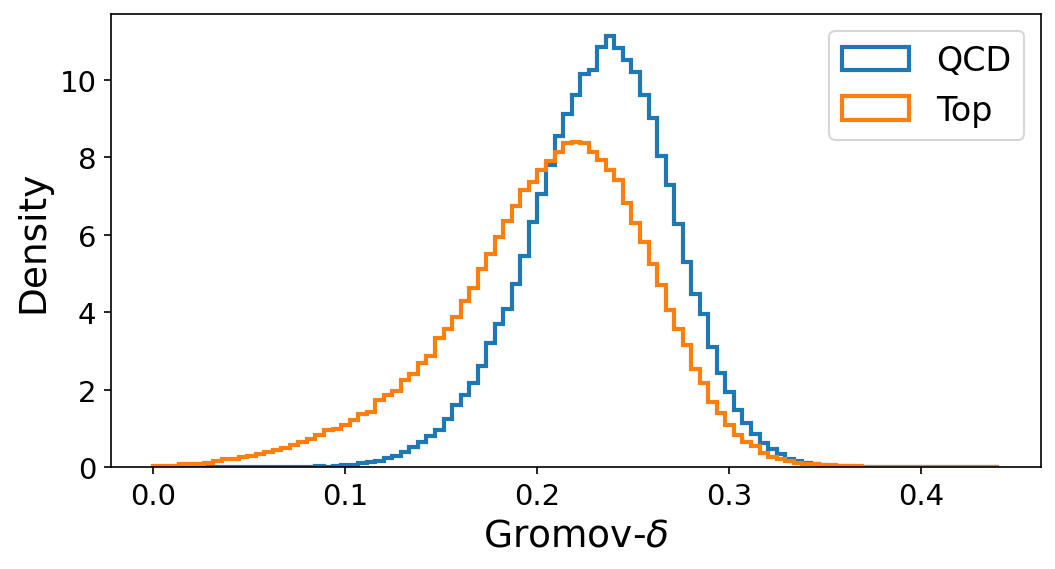}
    \vspace{-0.5cm}
    \caption{Distribution of per-jet Gromov-\(\delta\) hyperbolicity for QCD and top jets. Top jets have lower Gromov-\(\delta\) hyperbolicity, higher hierarchical structure, compared to QCD jets.}
    \label{fig:top_gromov_hist}
    \vspace{-1em}

\end{wrapfigure}

For each jet in the test dataset (400k total), we calculate the Gromov-\(\delta\) hyperbolicity on the preprocessed particle features without any zero-padding. Recall that lower values of Gromov-\(\delta\) hyperbolicity imply higher degrees of hierarchical structure. 
The distribution of Gromov-\(\delta\) hyperbolicity for top and QCD jets is shown in \Cref{fig:top_gromov_hist}. 
We find top jets display higher degrees of hierarchical structure in distribution compared to QCD jets.

We analyze jet-level features relative to Gromov-\(\delta\) hyperbolicity for \(p_T\), \(\eta\), and the number of constituents, with results presented in \Cref{sec:gromov_appendix} (\Cref{fig:gromov_jet_features}). Among these, we find the strongest correlation between hyperbolicity and the number of jet constituents; however, it remains unclear whether this relationship is an inherent feature or a byproduct of the Gromov-\(\delta\) hyperbolicity calculation. In \Cref{sec:gromov_appendix}, we compare the predicted classification probabilities for top and QCD jets relative to their Gromov-\(\delta\) hyperbolicity (\Cref{fig:prob_vs_gromov}) and observe a correlation between higher Gromov-\(\delta\) hyperbolicity (lower hierarchical structure) and misclassification, particularly for top jets.

\begin{figure}[h!]
    \centering
        \includegraphics[width=\linewidth]{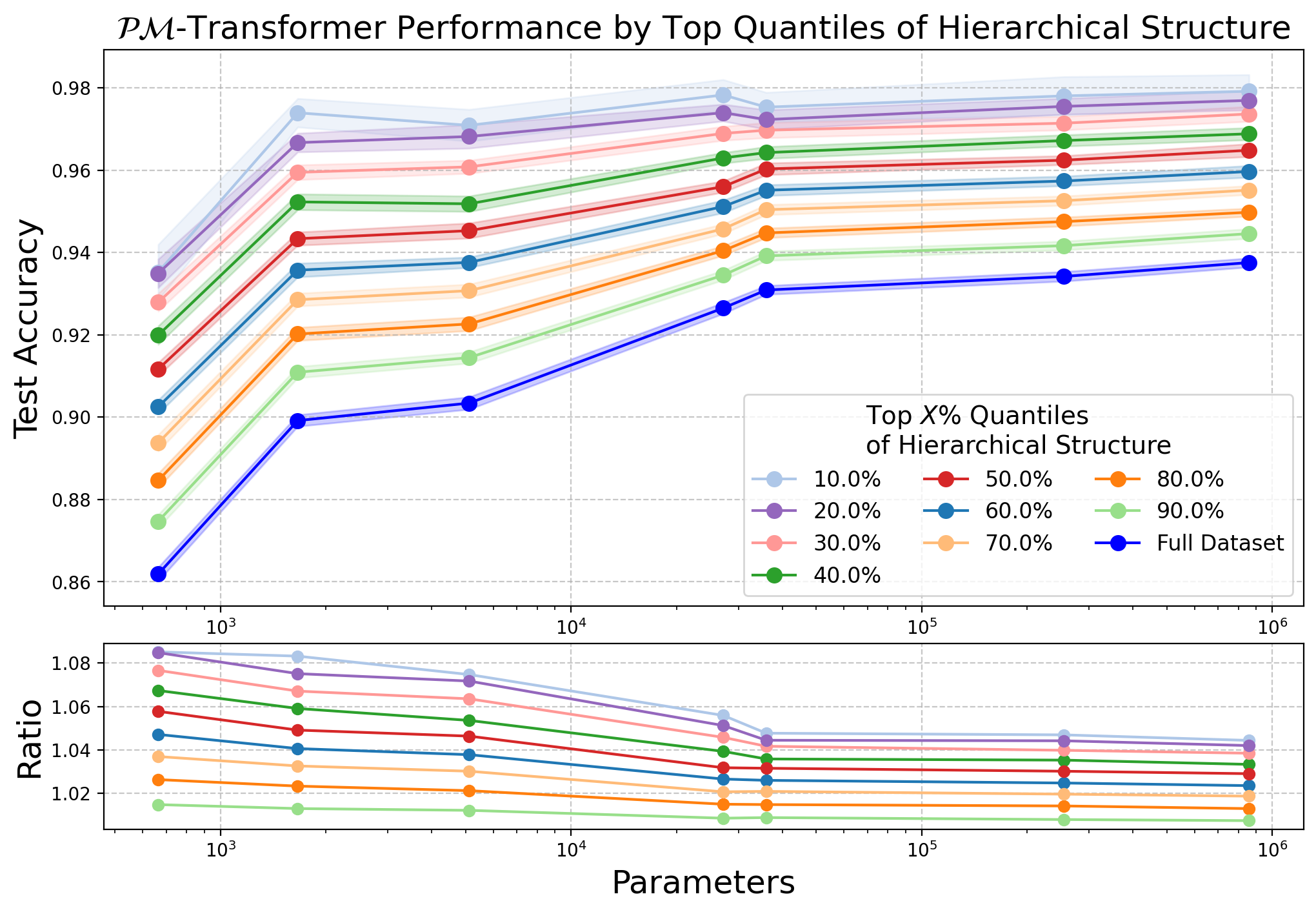}
        \caption{\(\mathcal P \mathcal M\)-Transformer test dataset performance plotted against the number of model parameters, shown for several top quantiles of hierarchical structure. Increasing the dataset’s hierarchical structure (by selecting smaller top quantiles) consistently yields performance gains across the entire parameter range tested. }
        \label{fig:gromov_large_filter}
    
\end{figure}

We analyze model performance in relation to Gromov-\(\delta\) hyperbolicity by evaluating test dataset performance after filtering jets based on different percentiles quantiles of their hierarchical structure. Specifically, we resample the test dataset, taking the top $X$\% quantile of the dataset according to the estimated hierarchical structure (the inverse of the calculated Gromov-\(\delta\) hyperbolicity). To avoid further biasing, after selecting the top \(X\%\) quantile, we resample the selected data to ensure an equal distribution of top and QCD jets in the filtered test dataset. We plot model performance relative to hierarchical structure quantiles in increments of 10\%, starting with the full dataset (400k jets) and progressing to the top 10\% most hierarchical jets (40k jets) in \Cref{fig:gromov_large_filter}. We include the same plot for the bottom quantiles in \Cref{sec:gromov_appendix}  (\Cref{fig:gromov_large_removed}) to showcase model performance on low data samples with low hierarchical structure.
We present results according to the accuracy as the rejection is limited by the shrinking dataset size.

We find that increasing the hierarchical structure of the dataset consistently corresponds to improved performance. This aligns with our preliminary results from the \(\mathcal{P} \mathcal{M}\)-MLP, which highlighted that highly hierarchical processes, as predicted by the Gromov-\(\delta\) hyperbolicity, benefit from hyperbolic representations.
Additionally, these performance gains are exaggerated for low-parameter models. For the smallest model (664 parameters), we see a 2.5\% and 5\% increases in performance on the top 80\% and 60\% quantiles of the dataset, respectively. These large gains suggests that \(\mathcal P \mathcal M\) representations could serve as a mechanism to shrink model parameters when processing hierarchical signals. 

The consistent correlation between increasing hierarchical structure and improved model performance across the full range of model parameters examined in \Cref{fig:gromov_large_filter} clearly demonstrates that hierarchical structure is a key factor in understanding and optimizing the processing of particle jets and, broadly, hierarchical data. This suggests that for hierarchical datasets, quantifying the degree of hierarchical structure at a per-sample level is crucial for understanding model performance across the range of samples in the dataset.

\section{Conclusion}
\label{sec:conclusions}

In this paper, we employ \(\mathcal{P}\mathcal{M}\) representations as a flexible tool for modeling hierarchical data and applied them to the challenging task of particle jet classification. We presented models ranging from small, MLP-based architectures to complex transformer models, all adapted to exploit \(\mathcal{P}\mathcal{M}\) representations. Our findings show that for hierarchical processes, such as those found in jet physics, \(\mathcal{P}\mathcal{M}\)-MLP models can achieve notable performance gains, especially in the regime of smaller, more parameter-efficient models.

Building on this foundation, we developed novel transformer architectures tailored to \(\mathcal{P}\mathcal{M}\) spaces. Benchmarking these models on jet tagging datasets revealed a clear correlation between hierarchical structure, as quantified by Gromov-\(\delta\) hyperbolicity, and model performance. We see the potential for the Gromov-\(\delta\) hyperbolicity as a useful analytic metric when dealing with hierarchical datasets, such as particle jets, as it provides meaningful insights into their underlying structure. 
% While our current \(\mathcal{P}\mathcal{M}\)-Transformer implementations do not surpass the performance of fully Euclidean counterparts or state-of-the-art models, this result motivates the exploration of further specialized architectures that can better leverage multiple geometric representations within the \(\mathcal{P}\mathcal{M}\) framework.
Our current \(\mathcal{P}\mathcal{M}\)-Transformer implementations achieve competitive performance relative to fully Euclidean counterparts and state-of-the-art models, highlighting the potential of the \(\mathcal{P}\mathcal{M}\) framework. These results inspire further exploration of specialized architectures to more effectively harness the benefits of multiple geometric representations.

Looking ahead, this work sets the stage for more advanced designs of \(\mathcal{P}\mathcal{M}\) models and for extending their use beyond classification to learning embeddings directly in \(\mathcal P \mathcal M\) representations. By demonstrating that \(\mathcal{P}\mathcal{M}\) representations can effectively capture the hierarchical relationships central to high-energy physics, we open the door to broader applications across diverse scientific fields. Ultimately, \(\mathcal{P}\mathcal{M}\) representations and the use of Gromov-\(\delta\) hyperbolicity as a descriptor offer a powerful lens through which to understand and analyze complex hierarchical datasets—whether they originate from particle physics, biology, social networks, or the Universe itself.

\section*{Acknowledgements}
This work is supported by the National Science Foundation under Cooperative Agreement PHY-2019786 (The NSF AI Institute for Artificial Intelligence and Fundamental Interactions, \href{http://iaifi.org/}{http://iaifi.org/}) and by the Paul E. Gray (1954) UROP Fund at MIT.
Computations in this paper were run on the FASRC Cannon cluster supported by the FAS Division of Science Research Computing Group at Harvard University. 

\bibliographystyle{plain}
\bibliography{main}
\appendix

\section{\(\mathcal P \mathcal M\)-MLP Particle-Level Embedding Visualizations}
\label{sec:plvl_embed}
\subsection{\(H \to 4q\) Embeddings}
\begin{figure}[H]
    \centering
    % First row, first column
    \begin{subfigure}[b]{0.45\textwidth}
        \centering
        \includegraphics[width=\textwidth]{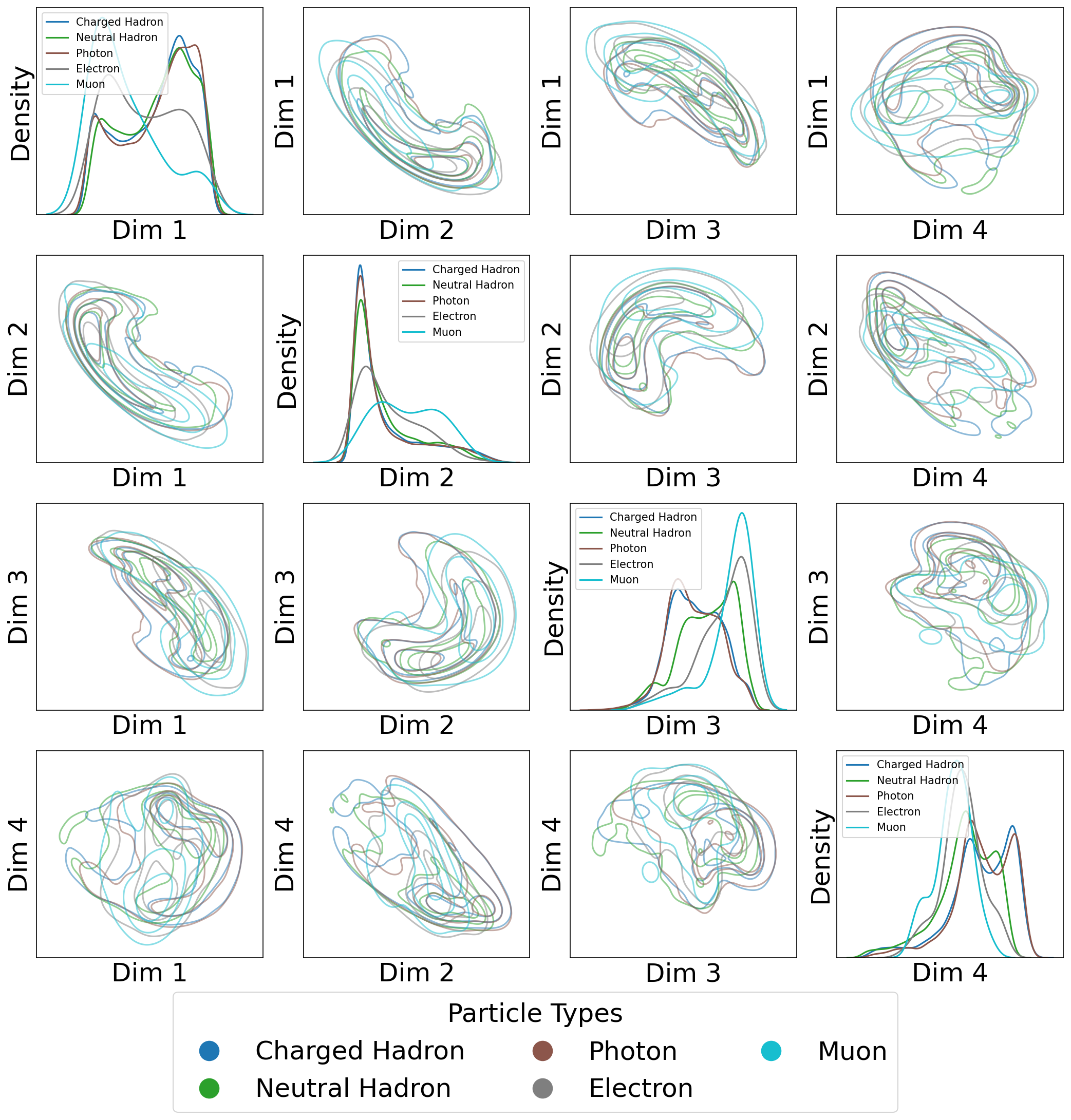}
        \caption{\(\mathcal T_0 \mathbb H^4\) corner plot colored by particle type}
        \label{fig:h4q_h4_kde}
    \end{subfigure}
    % First row, second column
    \begin{subfigure}[b]{0.45\textwidth}
        \centering
        \includegraphics[width=\textwidth]{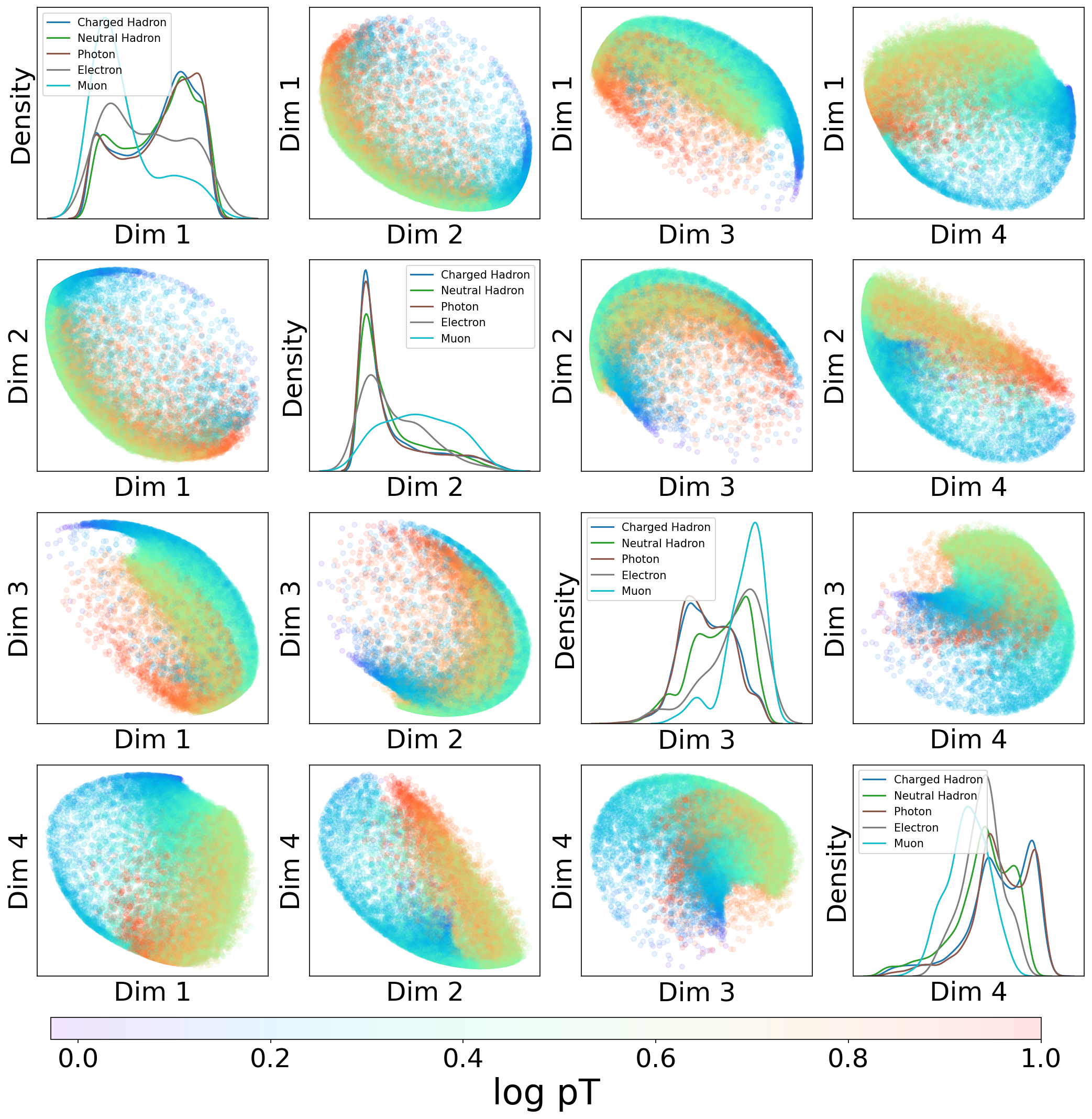}
        \caption{\(\mathcal T_0\mathbb H^4\) corner plot colored by $\log p_T$}
        \label{fig:h4q_h4_scatter}
    \end{subfigure}
    \\ % Line break to go to the next row
    % Second row, first column
    \begin{subfigure}[b]{0.45\textwidth}
        \centering
        \includegraphics[width=\textwidth]{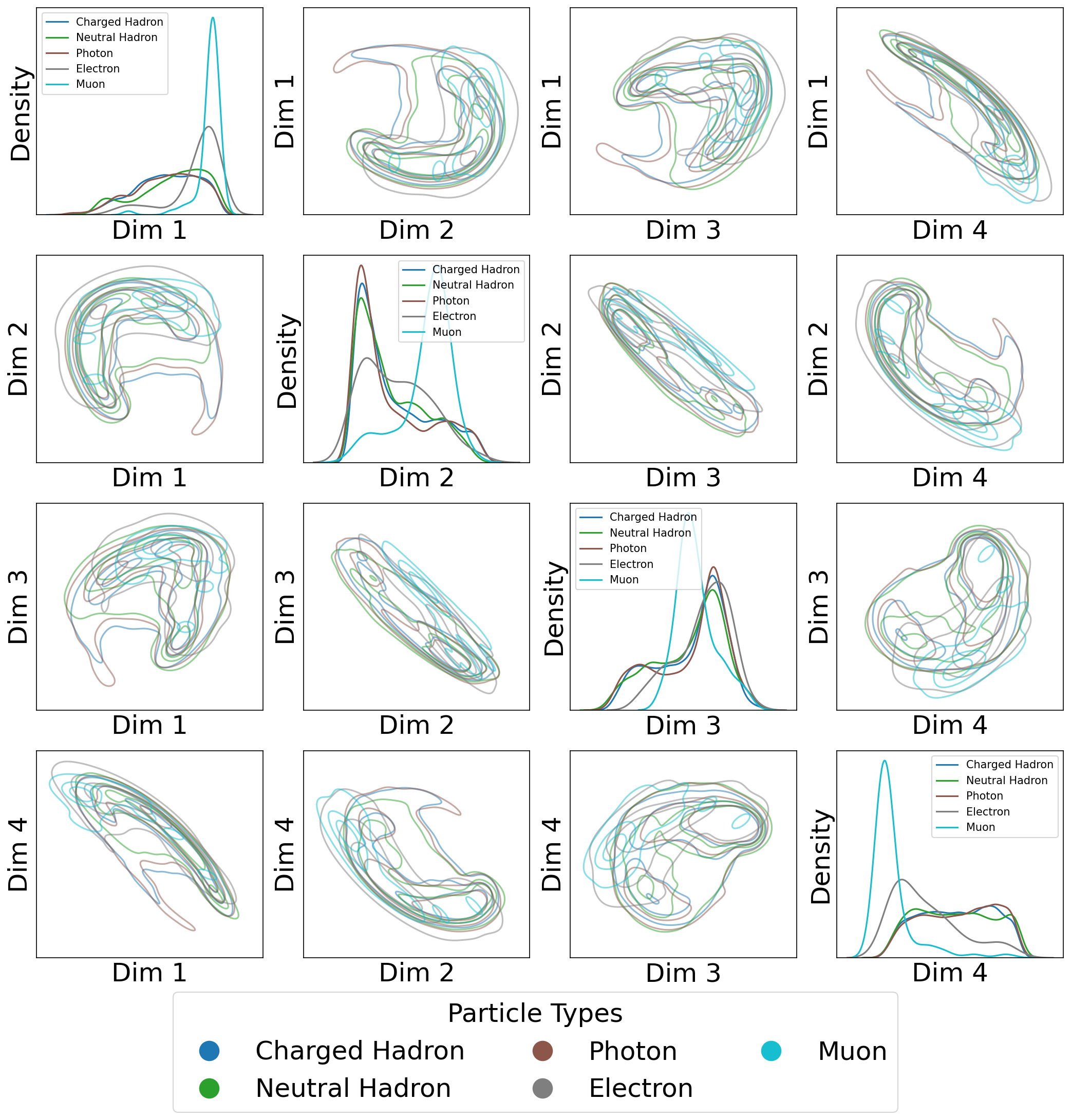}
        \caption{\(\mathbb R^4\) corner plot colored by particle type}
        \label{fig:h4q_r4_kde}
    \end{subfigure}
    % Second row, second column
    \begin{subfigure}[b]{0.45\textwidth}
        \centering
        \includegraphics[width=\textwidth]{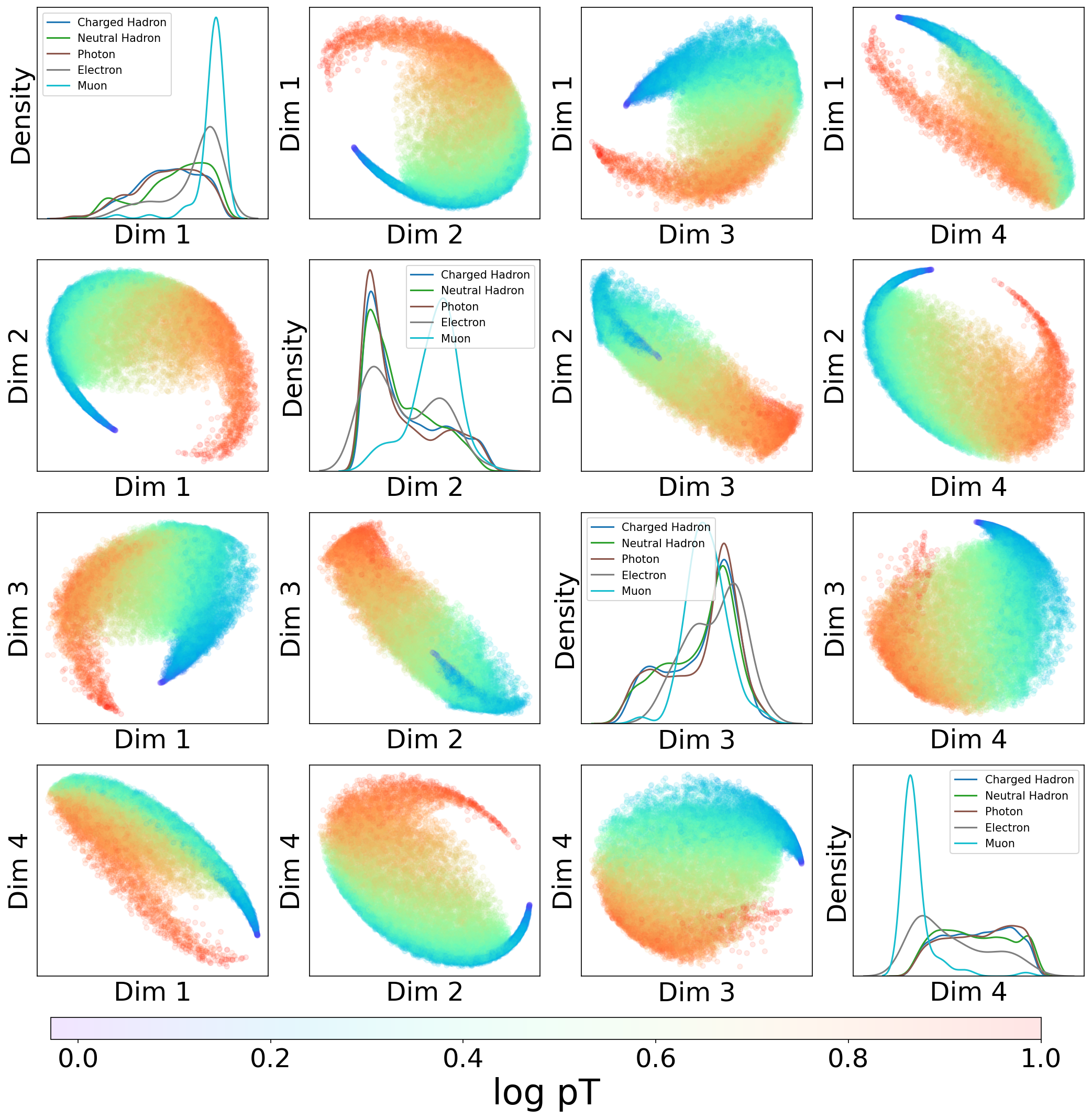}
        \caption{\(\mathbb R^4\) corner plot colored by $\log p_T$}
        \label{fig:h4q_r4_scatter}
    \end{subfigure}
    
    \caption{Corner plots for the 4D \(\mathcal P \mathcal M\)-MLP models for the \(H \to 4q\) process employing \(\mathbb R^4\) and \(\mathbb H^4\) geometries. For \(\mathbb H^4\), we plot results in the tangent space \(\mathcal T_0 \mathbb H^4\). We show corner plots for both particle-level embedding geometries for the constituents of 10k jets from the test dataset. We show the particle-level embeddings colored by both particle-type and \(\log p_T\), normalized to max 1.}
    \label{fig:h4q_corner}
\end{figure}

\subsection{\(t \to bqq'\) Embeddings}
\begin{figure}[H]
    \centering
    % First row, first column
    \begin{subfigure}[b]{0.45\textwidth}
        \centering
        \includegraphics[width=\textwidth]{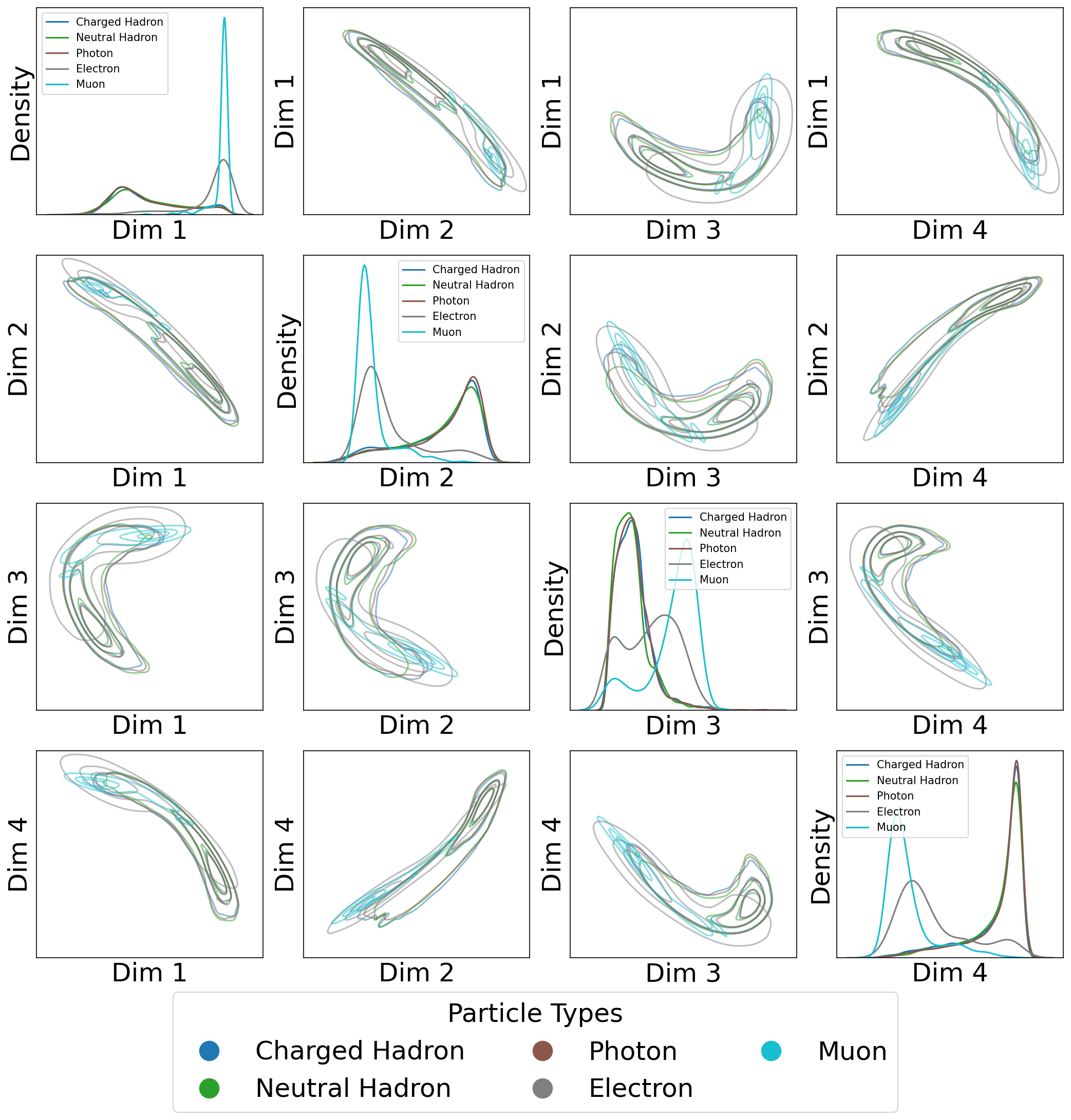}
        \caption{\(\mathcal T_0 \mathbb H^4\) corner plot colored by particle type}
        \label{fig:tbqq_h4_kde}
    \end{subfigure}
    % First row, second column
    \begin{subfigure}[b]{0.45\textwidth}
        \centering
        \includegraphics[width=\textwidth]{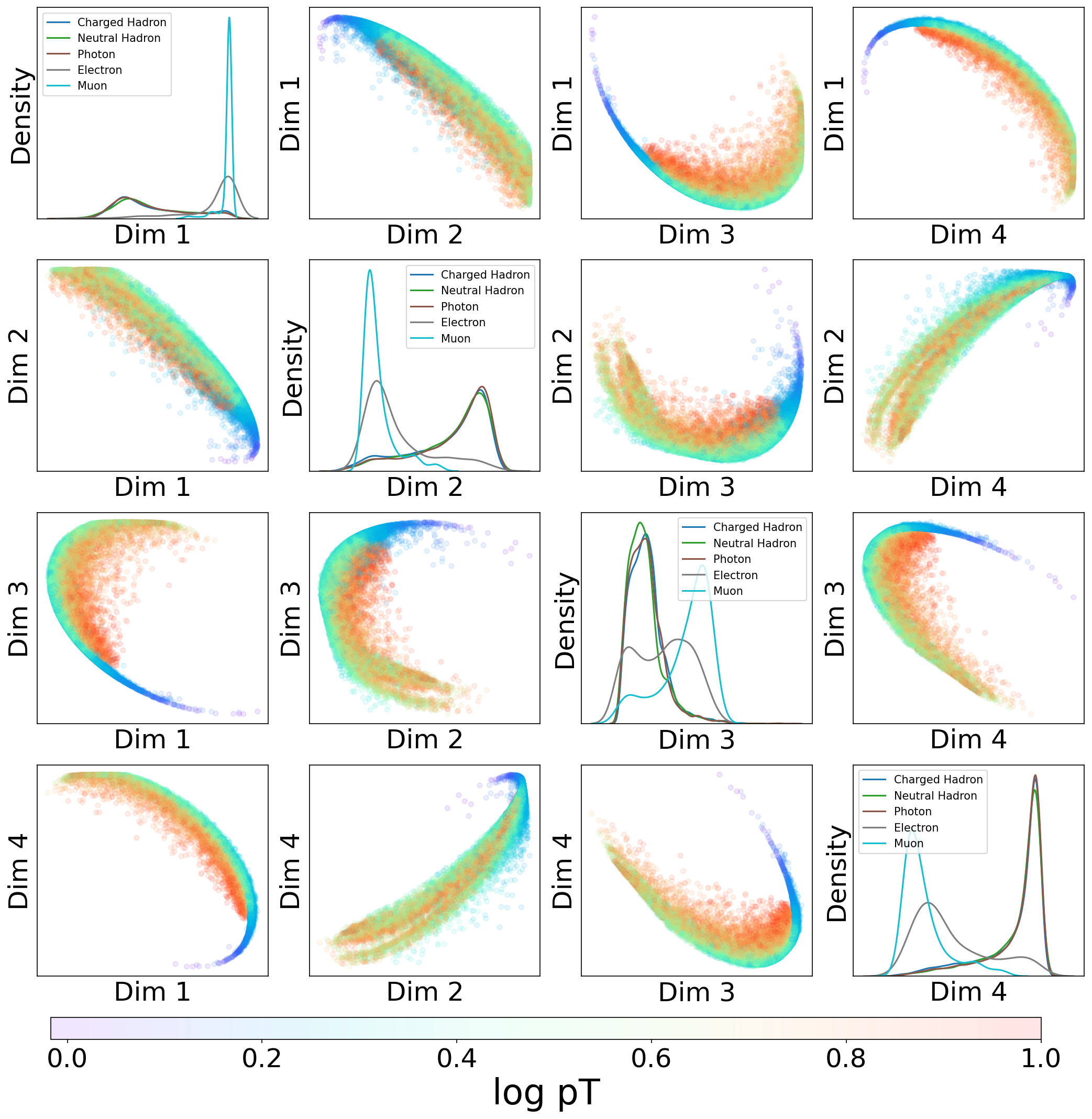}
        \caption{\(\mathcal T_0\mathbb H^4\) corner plot colored by $\log p_T$}
        \label{fig:tbqq_h4_scatter}
    \end{subfigure}
    \\ % Line break to go to the next row
    % Second row, first column
    \begin{subfigure}[b]{0.45\textwidth}
        \centering
        \includegraphics[width=\textwidth]{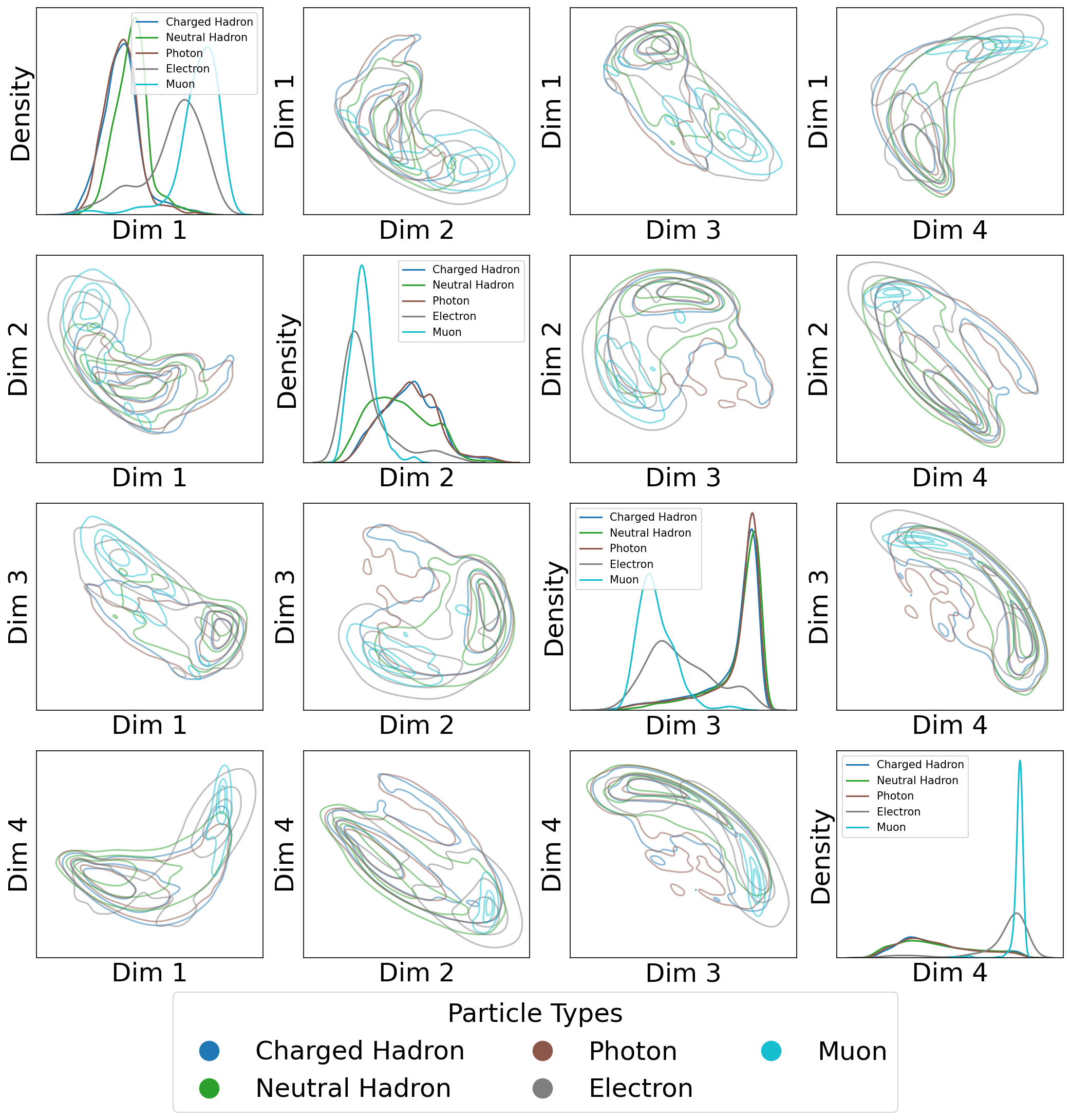}
        \caption{\(\mathbb R^4\) corner plot colored by particle type}
        \label{fig:tbqq_r4_kde}
    \end{subfigure}
    % Second row, second column
    \begin{subfigure}[b]{0.45\textwidth}
        \centering
        \includegraphics[width=\textwidth]{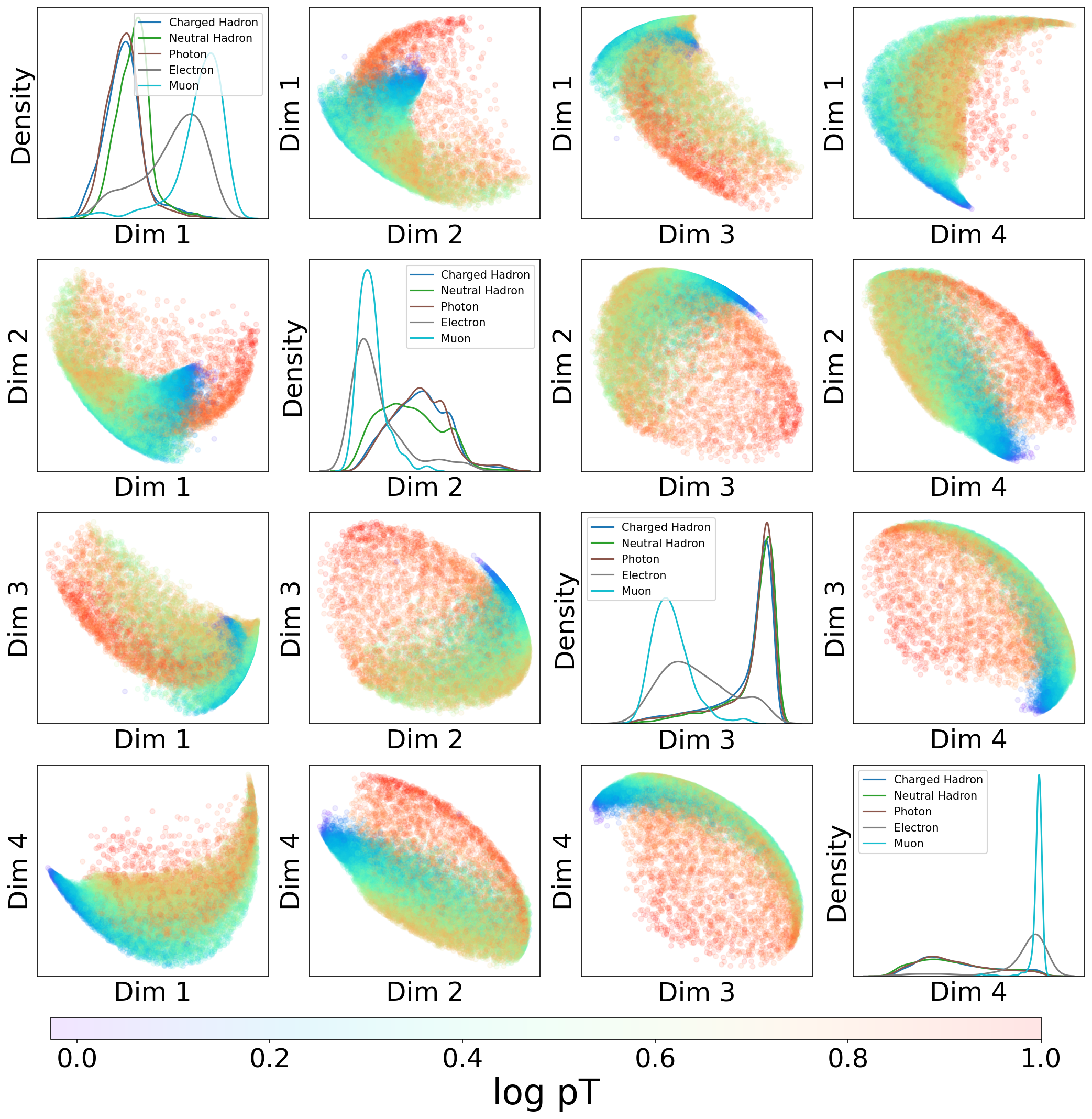}
        \caption{\(\mathbb R^4\) corner plot colored by $\log p_T$}
        \label{fig:tbqq_r4_scatter}
    \end{subfigure}
    
    \caption{Corner plots for the 4D \(\mathcal P \mathcal M\)-MLP models for the \(t \to bqq'\) process employing \(\mathbb R^4\) and \(\mathbb H^4\) geometries. For \(\mathbb H^4\), we plot results in the tangent space \(\mathcal T_0 \mathbb H^4\). We show corner plots for both particle-level embedding geometries for the constituents of 10k jets from the test dataset. We show the particle-level embeddings colored by both particle-type and \(\log p_T\), normalized to max 1.}
    \label{fig:tbqq_corner}
\end{figure}

\section{Top Tagging Gromov-\(\delta\) Hyperbolicity Resampling Additional Plots}
\label{sec:gromov_appendix}

\begin{figure}[H]
    \centering
    % First row with accuracy plots
    \begin{subfigure}[b]{0.49\linewidth}
        \centering
        \includegraphics[width=\linewidth]{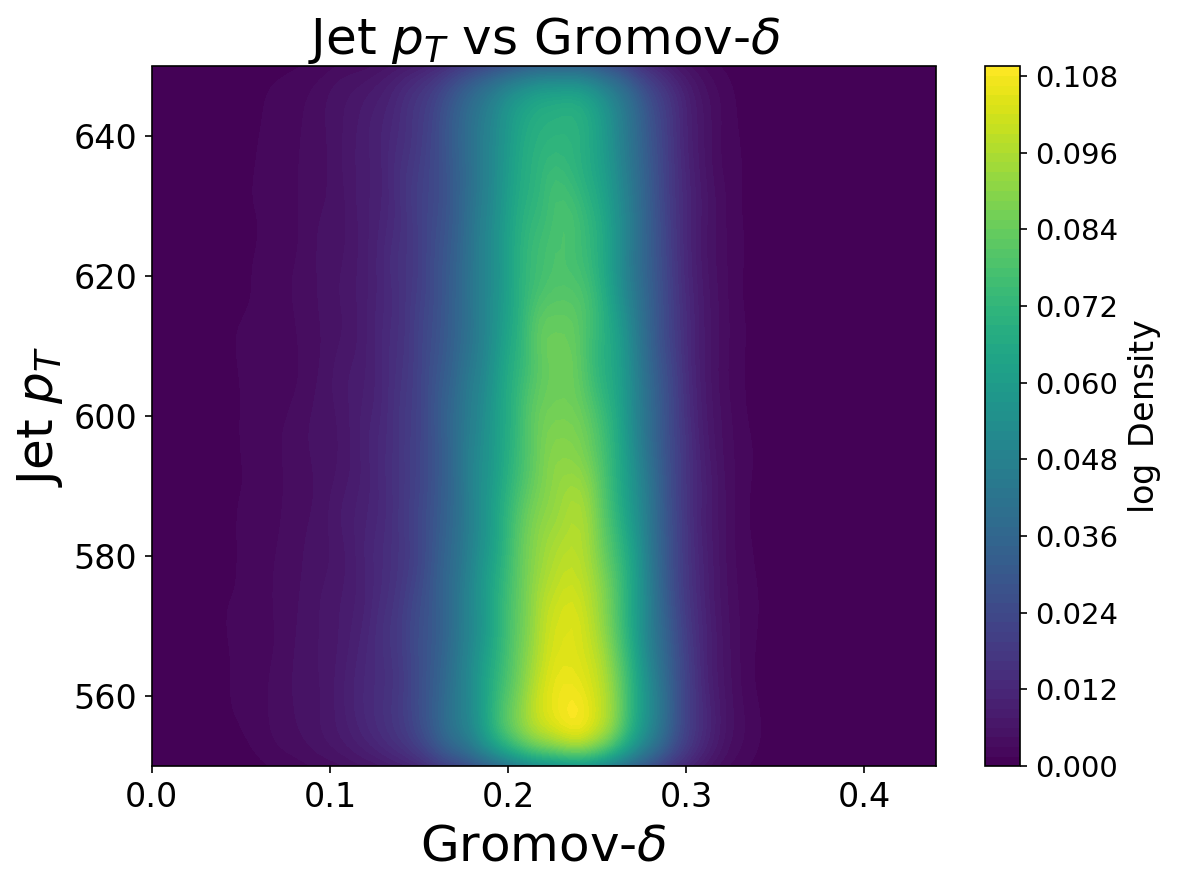}
        \caption{Jet \(p_T\) compared to calculated Gromov-\(\delta\) hyperbolicity}
        \label{fig:pt_vs_gromov}
        
    \end{subfigure} 
    \hfill
    \begin{subfigure}[b]{0.49\linewidth}
        \centering
        \includegraphics[width=\linewidth]{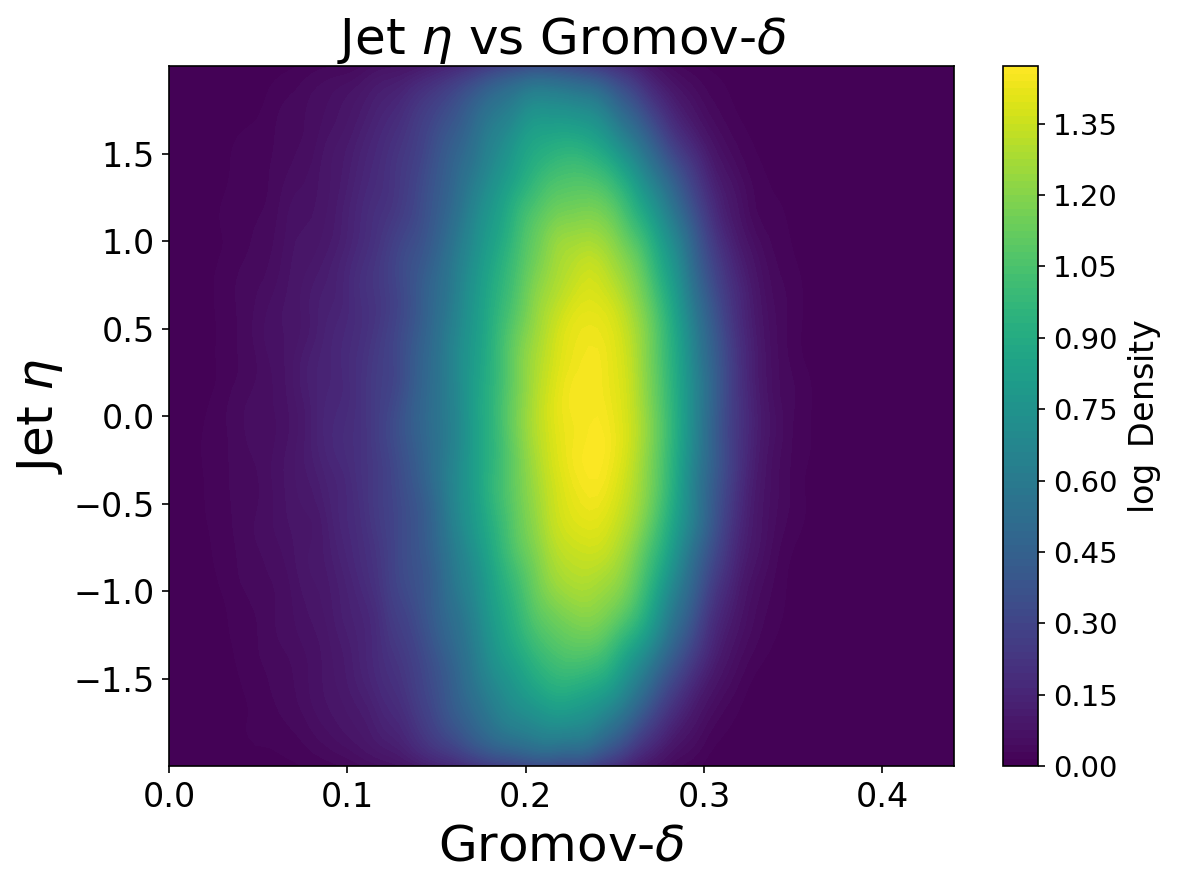}
        \caption{Jet \(\eta\) compared to calculated Gromov-\(\delta\) hyperbolicity}
        \label{fig:eta_vs_gromov}
    \end{subfigure} 
    \hfill
    \begin{subfigure}[b]{0.49\linewidth}
        \centering
        \includegraphics[width=\linewidth]{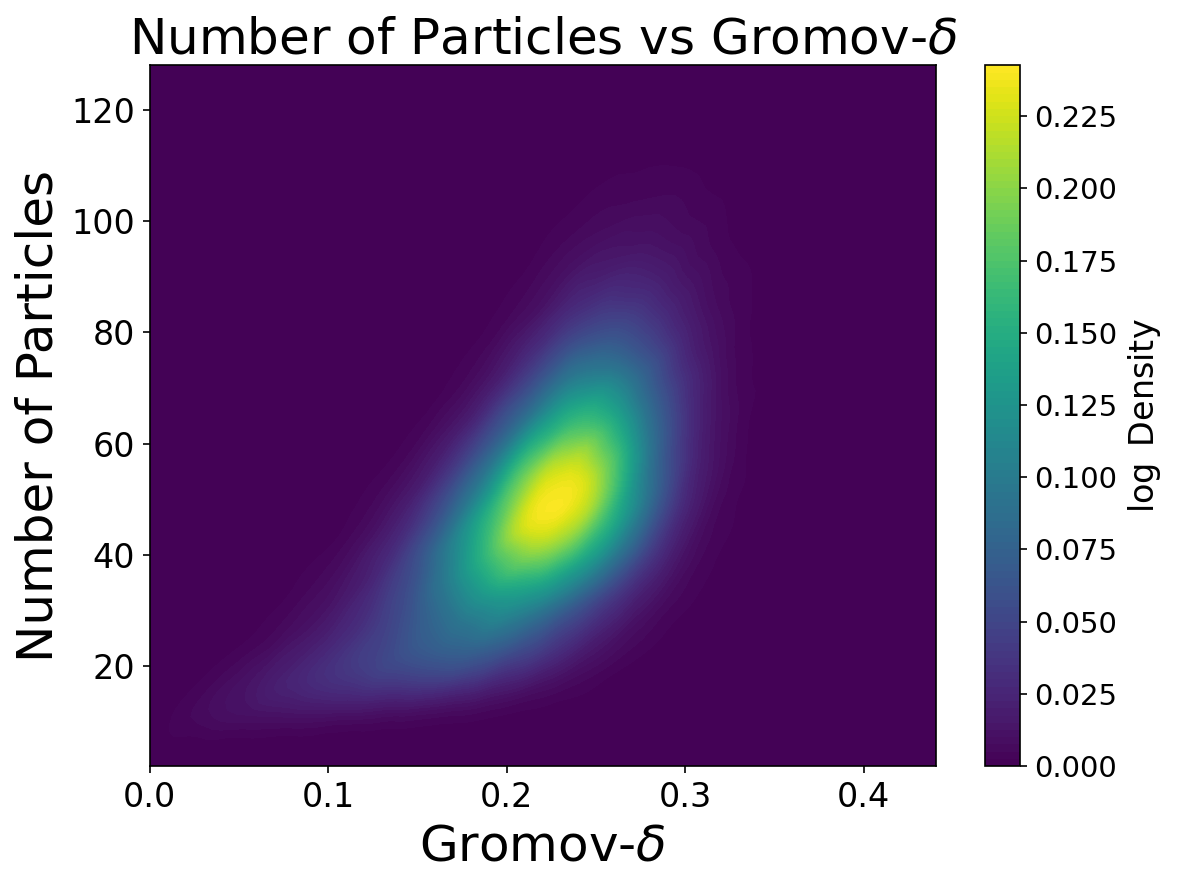}
        \caption{Number of particles compared to calculated Gromov-\(\delta\) hyperbolicity}
        \label{fig:num_parts_vs_gromov}
    \end{subfigure} 
    \caption{Comparison of jet-level features with Gromov-\(\delta\) hyperbolicity}
    \label{fig:gromov_jet_features}
   
\end{figure}

\begin{figure}[H]
    \centering
    % First row with accuracy plots
    \begin{subfigure}[b]{0.49\linewidth}
        \centering
        \includegraphics[width=\linewidth]{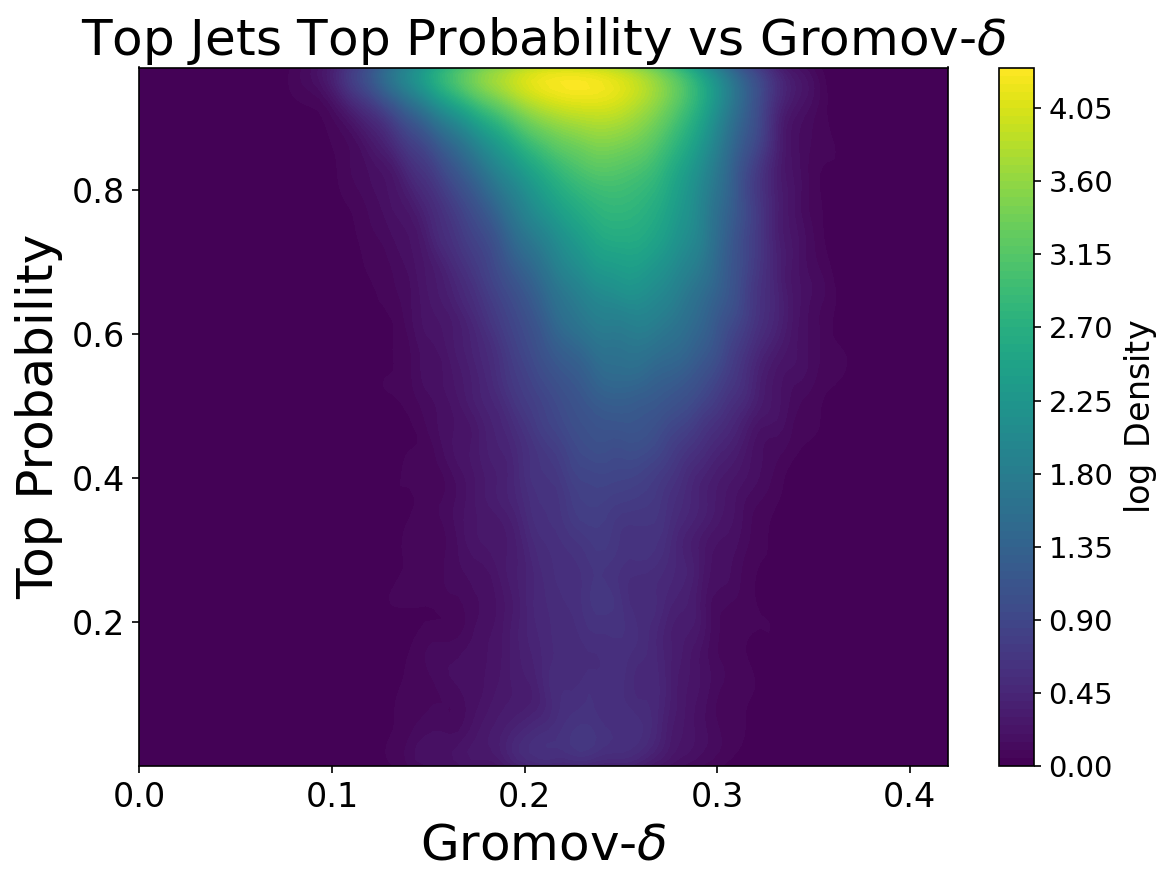}
        \caption{Top probability for top jets relative to calculated Gromov-\(\delta\) hyperbolicity}
        \label{fig:top_vs_gromov}
    \end{subfigure} 
    \hfill
    \begin{subfigure}[b]{0.49\linewidth}
        \centering
        \includegraphics[width=\linewidth]{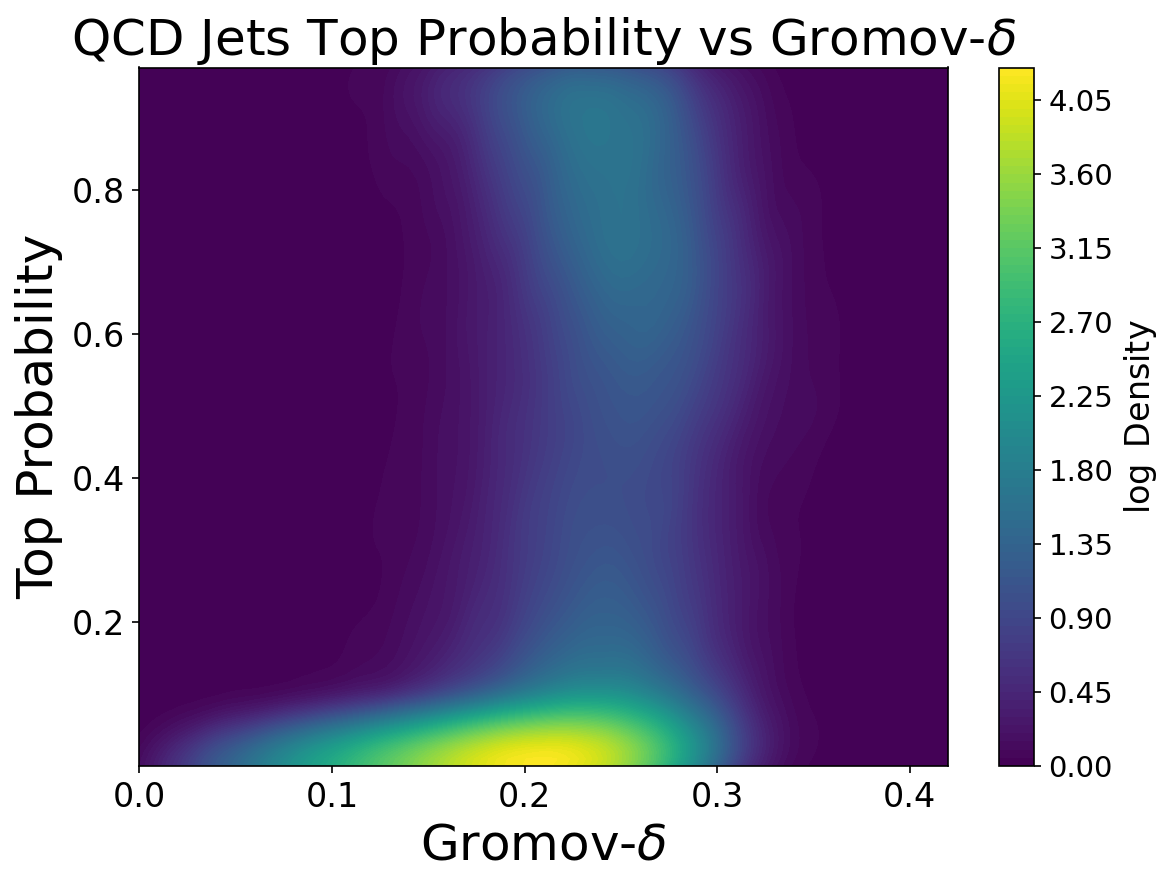}
        \caption{Top probability for QCD jets relative to calculated Gromov-\(\delta\) hyperbolicity}
        \label{fig:qcd_vs_gromov}
    \end{subfigure} 
    
    \label{fig:prob_vs_gromov}
    \caption{Classification probability as top jets for top jets (a) and QCD jets (b) relative to calculate Gromov-\(\delta\) hyperbolicity.}
   
\end{figure}

\begin{figure}[H]
    \centering
        \includegraphics[width=\linewidth]{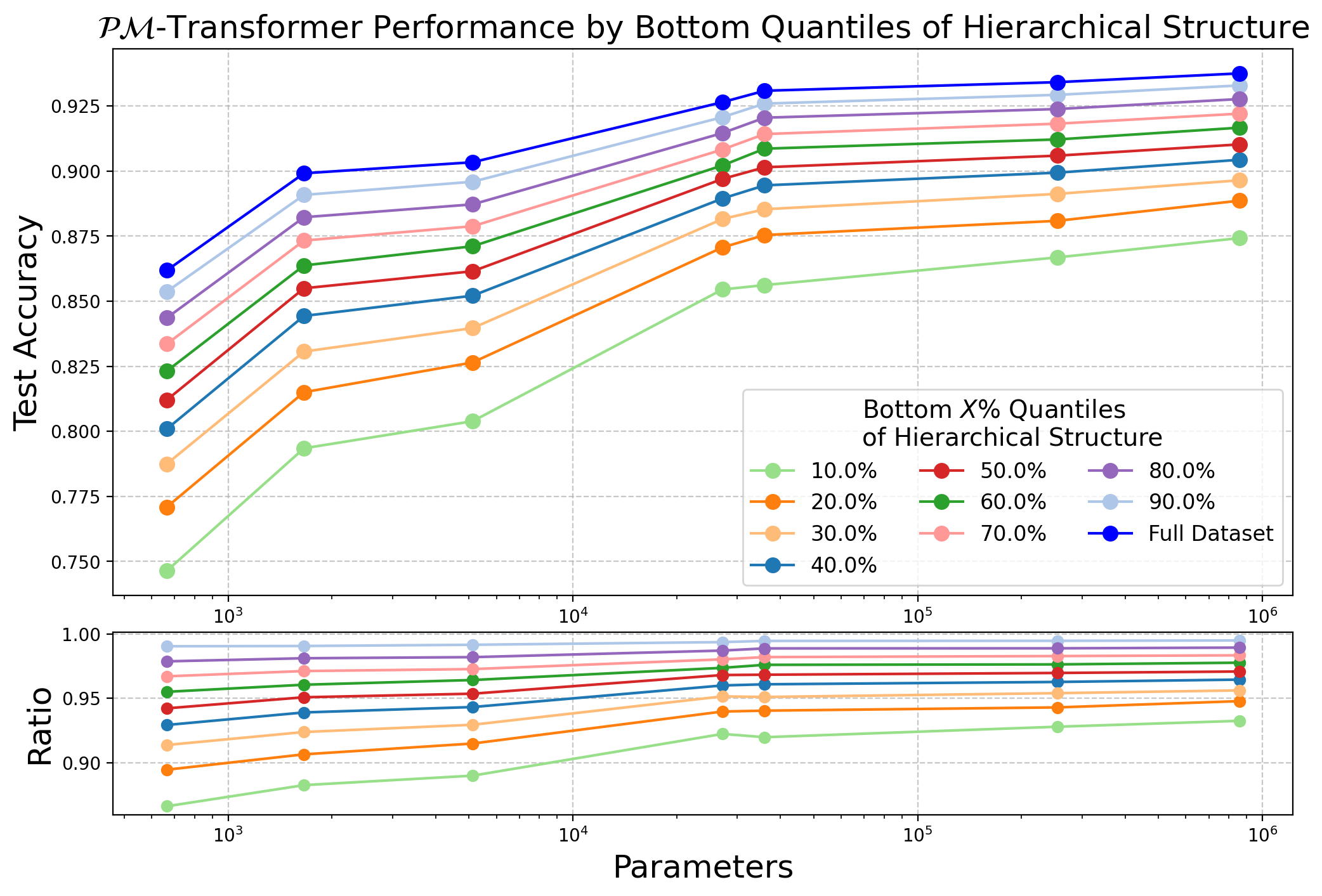}
        \caption{We present the accuracy of the \(\mathcal{P} \mathcal{M}\)-Transformer model on datasets filtered based on Gromov-\(\delta\) hyperbolicity. Specifically, the 80\% line represents the model's accuracy on data where the top 80\% of hyperbolicity values (i.e., the bottom 20\% of hyperbolicity) have been removed.}
        
    \label{fig:gromov_large_removed}
    
\end{figure}

\end{document}